\documentclass[12pt]{amsart}
\usepackage[left=1in,top=1in,right=1in,bottom=1in]{geometry}
\usepackage[utf8]{inputenc}
\usepackage{amsmath,natbib}
\usepackage{amsthm}
\usepackage{amsfonts}
\usepackage{mathtools}
\usepackage{amssymb}
\usepackage{graphicx}
\usepackage{listings}
\usepackage{color}
\usepackage{verbatim}
\usepackage{amsthm}
\usepackage{xcolor}
\usepackage[T1]{fontenc} 
\usepackage{fourier}
\usepackage{float}
\usepackage{pgfplotstable}
\usepackage{booktabs}
\usepackage{colortbl}
\usepackage{multirow}
\usepackage{array}
\usepackage{longtable}
\usepackage{tabu}
\usepackage{rotating}
\usepackage{lscape}
\usepackage{afterpage}
\usepackage{setspace}
\usepackage{multirow}
\usepackage{subfigure}

\pdfoutput=1

\definecolor{lightgray}{rgb}{0.95,0.95,0.95}
\numberwithin{equation}{section} 
\numberwithin{figure}{section} 
\numberwithin{table}{section} 

\newcommand\norm[1]{\left\lVert#1\right\rVert}

\begin{document}




\title{Optimal ETF Selection for Passive Investing}

\author{David Puelz, Carlos M. Carvalho and P. Richard Hahn}

%
%
%
%
\begin{abstract} 
This paper considers the problem of isolating a small number of exchange traded funds (ETFs) that suffice to capture the fundamental dimensions of variation in U.S. financial markets. First, the data is fit to a vector-valued Bayesian regression model, which is a matrix-variate generalization of the well known stochastic search variable selection (SSVS) of \cite{GeorgeandMcCulloch}. ETF selection is then performed using the ``decoupled shrinkage and selection'' procedure described in \cite{HahnCarvalho}, adapted in two ways: to the vector-response setting and to incorporate stochastic covariates. The selected set of ETFs is obtained under a number of different penalty and modeling choices. Optimal portfolios are constructed from selected ETFs by maximizing the Sharpe ratio posterior mean, and they are compared to the (unknown) optimal portfolio based on the full Bayesian model.  We compare our selection results to popular ETF advisor Wealthfront.com. Additionally, we consider selecting ETFs by modeling a large set of mutual funds.

\end{abstract}

\maketitle


\thispagestyle{empty}


%
%

\section{Introduction}
Exchange traded funds (ETFs) have emerged in recent years as a low-fee way for individuals to invest in the stock market. The growth of ETF popularity the past 20 years stemmed from investors' desire to participate passively in the returns of stocks in the overall market.  The first ETF began trading in January 1993 and was called the S\&P 500 Depository Receipt, also known as SPDR.  Since then, the size of the ETF market has grown to over \$1 trillion, and SPDR, a company derived from State Street Global Advisors, is the world's second largest ETF provider with assets of nearly \$350 billion.  ETF investing spans a large variety of asset classes, with funds holding currencies, foreign equity, bonds, real estate, and commodities.  The explosive growth of the ETF industry underscores the desire of the average investor to hold a diversified and broadly exposed portfolio for a cheap fee.

Although there are many fewer ETFs than there are generic tradable assets, an individual investor who has decided to invest entirely in ETFs still has decisions to make. Should one hold a variety of specialty ETFs, such as funds with a real-estate or biotech focus? Or is it adequate to hold a single broad-spectrum ``market" fund, such as the Russell 4000, which holds positions in thousands of individual stocks?  In this paper, we perform variable selection on ETFs to reduce the options a (long-term) investor faces to just a handful of distinct funds. 

To determine a small subset of ETFs most suitable for individual investing, our strategy will be to isolate those ETFs which capture the vast majority of variability in the stock market. Specifically, our analysis focuses on eight ``financial anomalies" from the asset pricing literature.  These assets are themselves formed as linear combinations of many individual stocks (according to an established recipe which involves pre-sorting the stocks by various criteria). Our working premise is that these eight assets represent a desirable cross-section of market risk for investors to be exposed to, see \cite{FF3} and \cite{FF5}. Granting this premise, an investor still cannot invest in these eight factors directly for practical reasons; trading costs from thousands of buy-sell transactions prohibit this strategy (although several mutual fund providers such as Dimensional Fund Advisors sell products that attempt to mimic these theoretical strategies). For this reason, we refer to our response vector as the ``unattainable or target assets."  With this as background, our goal is simply to find a small number of ETFs that replicate the covariance structure of the unattainable assets to a reasonable practical tolerance. Once these ETFs are selected, various portfolio optimization strategies can be implemented. We compare the performance of these portfolios to the inferred performance of the (unknown) optimal portfolio implied by our statistical model.

Methodologically, our analysis combines and extends two previous techniques. First, we extend the decision-theoretic variable selection (DSS) approach of \cite{HahnCarvalho} to the vector-valued response setting. The DSS approach consists of two phases, a model fitting phase and a variable selection phase. In the model-fitting phase, we adapt and extend the stochastic search variable selection (SSVS) \citep{GeorgeandMcCulloch,BrownVannucci98} for a vector-valued response. This model differs from a naive application of SSVS to a vector-valued response in that variable inclusion is determined simultaneously across the individual univariate regressions; that is, the variable either appears in all of the regressions or none of them.  Also, in the selection phase, we consider a stochastic design matrix; Hahn and Carvalho (2015) consider only a fixed-design utility function which draws a natural connection to model selection in gaussian graphical models as we will seek to explore the conditional independence relationships between the ETFs and the target assets \citep{Jonesetal2005,WangCarvalho2011,Wang2015}. This modification is important in the context of investing, because the future returns of the ETFs are unknown at the time of selection. 

\subsection{Previous ETF research}
Recent research has focused on evaluating ETFs as single investments.  \cite{Poterba} examine the operation of ETFs from a tax efficiency perspective.  They conclude that ETFs are more tax efficient than equity mutual funds by noting that taxable gains on ETFs are smaller than comparable mutual funds, suggesting they are a reasonable low-cost investment for taxable investors.  \cite{Agapovan} compares passive, index tracking mutual funds and ETFs.  She examines fund flows using a pooled OLS model and finds that ETFs are almost perfect substitutes for passive mutual funds.  \cite{DiLellio} look at whether published ETF trading strategies outperform the market.  They found many strategies outperform the S\&P 500 but with weak statistical significance.  Several other papers study investment characteristics of ETFs, including \cite{Huanga}, \cite{Shin}, \cite{Pennathur}, \cite{Ackert} and \cite{Kostovetsky}.  We contribute to this diverse body of research by proposing an investment methodology for the average investor using ETFs as the sole financial product. 

The construction of passive portfolios is a separate area of the literature but also relevant to our research.  These problems are typically framed in a variable selection framework in which regularization and optimization become important tools.  Index tracking is one approach to forming such a portfolio.  This is done by determining which subset of index components can be invested in while maintaining similar performance to the index - commonly known as index tracking.  \cite{Rockafellar} present a conditional value-at-risk (CVaR) constrained optimization and apply it to tracking the S\&P 100 index. \cite{Fastrich} consider penalized optimization to construct sparse optimal portfolios.  They review empirical performance of portfolios built from several penalties, including $l_{q}$-regularizers, and develop a new penalty that leads to high Sharpe ratio (risk-adjusted return) tracking portfolios.  In two separate papers, Wu et al. (2014) consider the special cases of $l_{1}$ and $l_{2}$ penalties in their optimization, known as the \textit{lasso} and \textit{elastic net}, respectively \citep{Wu, Wu2}.  They develop algorithms to solve a nonnegative optimization problem where the decision variables are the long-only weights on assets in a tracking portfolio.  In other words, they do not allow the short-selling of an asset. A Nonnegative Irrepresentable (NIR) condition is also shown to guarantee variable selection consistency.  Since picking a subset of countably many assets is the goal, exploring all possible combinations of assets can be undertaken with mixed-integer programming (MIP).  \cite{Canakgoz} develop an MIP approach to index tracking as well as enhanced indexation, where the objective is to outperform the index.  This approach includes transaction costs and linearization of the tracking portfolio returns for computational tractability.  Similarly, \cite{Chen} consider a robust MIP formulation by incorporating estimation error into the objective quantities.  They develop a fast algorithm by maximizing pairwise similarities between assets in the tracking portfolio and target index. \cite{Beasley} develop an evolutionary heuristic for the index tracking problem which incorporates transaction costs.  They consider a minimization problem involving the tracking error and excess return and discuss in sample and out of sample performance.
	
	Uncertainty is central to the problem at hand.  It rears its head through parameters in the statistical model we specify and the asset returns we use for estimation.  \cite{JandP} review Bayesian tools in finance used to deal with uncertainty.  They discuss a framework for evaluating predictive distributions of unknown returns and parameters and how one deals with financial quantities, such as the Sharpe ratio, from a Bayesian perspective.  \cite{PandV} survey recent literature focused on ``learning in financial markets."  The executive summary: acknowledging parameter uncertainty leads to easier interpretation of common models used in finance. This is the mantra of our approach. We take into account the unknown future by integrating over parameter and return uncertainty before the selection of ETFs is made.


\section{An ETF factor model of market covariation}
Our analysis revolves around eight financial ``anomalies" from the finance literature, which go by the names: size, value, market, direct profitability, investment, short-term reversal, long-term reversal, and momentum.  Each anomaly is a portfolio, constructed by cross-sectionally sorting stocks by various characteristics of a company and forming linear combinations based on these sorts.  For example, the value anomaly is constructed using the book-to-market (the value of a company ``on paper" divided by the market's perception of its value) ratio of a company.  A high ratio indicates the company's stock is a ``value stock" while a low ratio leads to a ``growth stock" assessment.  Essentially, the value anomaly is a portfolio built by going long stocks with high book-to-market ratio and shorting stocks with low book-to-market ratio.  For detailed definitions of the first five factors, see \cite{FF5}. The data we use in our analysis was obtained from Ken French's website\footnote{http://mba.tuck.dartmouth.edu/pages/faculty/ken.french/}. It is widely believed that these eight anomaly portfolios (or some subset of them) reflect all dimensions of independent variation in the stock market as reflected in \cite{FF3} and \cite{FF5}. 

Although these strategies cannot be readily implemented by the average investor, we might hope to determine a set of ETFs that recapitulates their covariance structure. To this end, we consider the 25 most highly traded (i.e., most liquid) equity funds from ETFdb.com.  For more recent data periods, we are able to increase our number of ETFs to 46.  Specifically, we use monthly ETF data from the Center for Research in Security Prices (CRSP) database from February 1992 through February 2015 \citep{CRSP}. 

In the next section, we lay out a regression model that ties these attainable assets (ETFs, which are easy to invest in) to the target assets (the eight anomalies, with returns we may readily observe, but not readily invest in).

\subsection{The regression model}
Arbitrage pricing theory (APT) \citep{Ross} expresses expected returns as a linear combination of systematic factors and sensitivity parameters:
\begin{align}
	\mathbb{E}[R_{j}] &= r_{f} + \beta_{1j}F_{1} + \cdots + \beta_{pj}F_{p}, 
\end{align}where $R_{j}$ denotes returns and $F_{p}$ represent (possibly unobservable) sources of undiversifiable risk --- the unavoidable risk inherent to putting one's money in the market. 
The theory derives its name from its assumption that all asset prices divergent from the model will be corrected by arbitrage.  

Our model will assume that we we may define our systematic factors in terms of ETFs. That is, given a set of target asset returns, $\{ R_{j} \}_{j=1}^{q}$, and ETFs, $\{ X_{i} \}_{i=1}^{p}$, we model the target returns as:
\begin{equation}
	R_{j} = \beta_{j1}X_{1} + \cdots + \beta_{jp}X_{p} + \epsilon_{j}, \;\;\;\; \epsilon_{j} \sim \textrm{N}(0,\sigma^{2}).
\end{equation}\label{model1} We argue that it is reasonable to fix the ETFs as factors in an APT model because there are many such funds and that they trade across multiple asset classes and markets.  
In this formulation, the right-hand side of model \ref{model1} represents the set of assets attainable for the average investor.  The left-hand side are unattainable but desirable assets --- the target assets.  The linear model provides a mapping between the attainable and unattainable spaces that can be rigorously studied.  The remaining challenge is to determine a small number of ETFs that simultaneously well-approximate all eight of the target returns.

For $T$ periods, our linear model can be expressed compactly as a matrix normal distribution \citep{dawid1981some}. Define the matrix of target assets as $\textbf{R} \in \mathbb{R}^{Txq}$ and the matrix of ETFs as $\textbf{X} \in \mathbb{R}^{Txp}$.  Additionally, let $\gamma \in \{0,1\}^{p}$ be a binary vector identifying a particular ETF model where the nonzero entries specify which ETFs are included. We write the model $M_{\gamma}$ as:
\begin{equation}\label{themodel}
\begin{split}
	M_{\gamma}: \hspace{2mm} \textbf{R} \sim \textrm{Matrix Normal}_{T,q}\left(\textbf{X}_{\gamma} \boldsymbol \beta_{\gamma}, \hspace{1mm} \sigma^{2}\mathbb{I}_{T \times T}, \hspace{1mm} \mathbb{I}_{q \times q}\right).
	\end{split}
\end{equation}
Note the row and column covariances are diagonal by the APT assumptions.


%

\section{Utility-based ETF selection}

Our analysis adapts the model selection approach described in \cite{HahnCarvalho}, who cast model selection as a means to an end.  They argue that if the goal is selection of a small subset of covariates, that desire should be reflected in a utility function rewarding sparsity (rather than via a prior).  They propose a \textit{DSS loss function (decoupled shrinkage and selection)} derived by integrating over the predictive and posterior distributions from standard model space sampling as in \cite{GeorgeandMcCulloch}.  Once integrating over posterior uncertainty, this loss function is used for covariate selection.  Assuming the design matrix and prediction points are identical and given by $\textbf{X}$, the DSS loss function is:

\begin{align} \label{DSS}
	\mathcal{L}(\gamma) = T^{-1} \| \textbf{X}\bar{\beta} - \textbf{X} \gamma \|_{2}^{2} + \lambda \|\gamma\|_{0},	
\end{align}where $\bar{\beta}$ is the posterior mean and $\gamma$ is the choice variable.  The loss function elegantly depends only on the posterior mean, $\bar{\beta}$.  Approximating the penalty term in \ref{DSS} with an L1-norm, the selection step amounts to solving the predictive loss minimization problem.

\begin{align} \label{DSS2}
	\beta_{\lambda} \coloneqq \arg \min_{\gamma} T^{-1} \| \textbf{X}\bar{\beta} - \textbf{X} \gamma \|_{2}^{2} + \lambda \|\gamma\|_{1},
\end{align}where $\beta_{\lambda}$ is sparse since the objective function is penalized.  Thus, the nonzero elements of $\beta_{\lambda}$ determine which covariates are selected.  Hahn and Carvalho discuss approaches for choosing a tuning parameter $\lambda$ along the solution path.  We use the two step approach of this paradigm in our analysis, outlined as:

\begin{enumerate}
	\item Model fitting step: Modeling the marginal ETF distribution and sampling the conditional model space via Bayesian conditioning,
	\item Selection step: Integrate over posterior uncertainty and determine a sparse selection of covariates.
\end{enumerate}

Many approaches can be used for the selection step including lasso optimization as in \ref{DSS2} or naive forward stepwise selection.  Regardless of the approach, concerns of overfitting are sidestepped by working with a denoised target, $\textbf{X}\bar{\beta}$. Given this pre-smoothed response, it is natural to think of the selection step as ``fitting the fit." \cite{RobSBIES}.

\subsection{Model fitting: The marginal and conditional distributions}

The future returns of the target assets and ETFs are unknown.  Acknowledging this uncertainty is important in the overall decision of which ETFs to select.  In fact, it is necessary for an honest ex ante selection of a subset of these assets.  We account for this by modeling the marginal distribution of the ETFs (denoted by the matrix $X$) via a latent factor model. The target assets are modeled conditionally via the APT model, and this procedure is described in the next subsection.  Using the compositional representation of the joint distribution:

\begin{align}
	p(x,r) = p(r \vert x)p(x).
\end{align}We specify the following model for the joint distribution:

\begin{align}
	\left[
	\begin{array}{c}
		\textbf{R} \\
		\textbf{X}
	\end{array}
	\right]
	\sim N(\mu,\Sigma),
\end{align}where $\Sigma$ has a block covariance structure:  

\begin{align}
\Sigma
= 
\left[
\begin{array}{c|c}
\beta^{T}\Sigma_{x}\beta + \Psi & (\Sigma_{x}\beta)^{T} \\
\hline
\Sigma_{x}\beta
 & 
\Sigma_{x} \\
\end{array}
\right].
\end{align}Notice that the upper right block is the marginal variance of $Y$ implied by the APT model. The lower right block is simply the marginal variance of $X$ we must additionally model.

We obtain posterior samples of $\Sigma$ by sampling the APT model parameters using a matrix-variate stochastic search algorithm (described below) and sampling the covariance of $X$ from a latent factor model where it is marginally normally distributed.  To reiterate our procedure is

\begin{itemize}
	\item $\Sigma_{x}$ is sampled from independent latent factor model,
	\item $\beta$ is sampled from matrix-variate MCMC,
	\item $\Psi$ is sampled from matrix-variate MCMC.
\end{itemize}

\subsubsection{Modeling the marginal distribution: A latent factor model}

We model ETFs via a latent factor model of the form:

\begin{equation}
	\begin{split}
		\textbf{X}_{t} &= \mu_{x} + \textbf{B}\textbf{f}_{t} + \textbf{v}_{t}
		\\
		&\textbf{v}_{t} \sim \text{N}(0,\mathbf{\Psi})
		\\
		&\textbf{f}_{t} \sim \text{N}(0,\textbf{I}_{k}),
		\\
		&\mu_{x} \sim \text{N}(0,\Phi)
	\end{split}
\end{equation}where $\Psi$ is assumed diagonal and the set of $k$ latent factors $f_{t}$ are independent.  The covariance of the ETFs is constrained by the factor decomposition and takes the form:

\begin{equation}
	\begin{split}
		\Sigma_{x} = \textbf{B}\textbf{B}^{T} + \Psi.
	\end{split}
\end{equation} To estimate this model, we use the R package {\tt bfa} from Jared Murray \citep{bfa}.  The software allows us to sample the marginal covariance as well as the marginal mean via a simple Gibbs step assuming a normal prior on $\mu_{x}$.

\subsubsection{Modeling the conditional distribution: A Matrix-variate stochastic search}

We model the conditional distribution, $\textbf{R} \vert \textbf{X}$, by developing a novel variable selection algorithm and sample parameters via Bayesian conditioning.  Recall that the conditional model is of the form \ref{themodel}, and we aim to explore the posterior on the model space, $\textbf{P}\left(M_{\gamma} \hspace{1mm} \vert \hspace{1mm} \textbf{R} \right)$.  This is a generalization of stochastic search variable selection from \cite{GeorgeandMcCulloch} in that our response is vector-valued instead of a single random variable.  Thus, the observed target asset data, $\textbf{R}$, is a matrix.

Similar to \cite{GeorgeandMcCulloch}, our algorithm explores the model space by calculating a Bayes factor for a paticular model $M_{\gamma}$.  Given that the response $\textbf{R}$ is matrix instead of a vector, we derive the Bayes factor as a product of vector response Bayes factors.  This is done by separating the marginal likelihood of the target assets as a product of distinct vector response marginal likelihoods for each of the target assets separately.  This derivation requires our priors to be independent across the target assets and is shown in the appendix.    Our approach is novel precisely because we have adapted SSVS to a matrix-variate response.  Note that we do not run standard SSVS on each target asset regression separately. Instead, we generalize \cite{GeorgeandMcCulloch} and require all covariates to be included or excluded from a model for all of the target assets \textit{simultaneaously}. 

The marginal likelihood requires priors for the parameters $\beta$ and $\sigma$ parameters in our model.  We use the well known g-prior that is standard for linear models because it permits an analytical solution for the marginal likelihood integral \citep{Z1,Z3,Liangetal08}.  

Our Gibbs sampling algorithm follows the standard stochastic search variable selection directly.  The aim is to scan through all possible covariates and determine which ones to include in the model, and this is how the algorithm explores the model space. At each substep of the MCMC where we are looking at an individual covariate within a specific model, we compute the probability of covariate's inclusion as a function of the model's prior probability and the Bayes factors:

\begin{equation*}
\begin{split}
	p_{i} = \frac{B_{a0} \textbf{P}\left(M_{\gamma_{a}}\right)}{B_{a0} \textbf{P}\left(M_{\gamma_{a}}\right) + B_{b0} \textbf{P}\left(M_{\gamma_{b}}\right)}.
\end{split}	
\end{equation*}The prior on the model space, $\textbf{P}\left(M_{\gamma}\right)$, can either be chosen to adjust for multiplicity or uniform - our results are robust to both specifications.  In this setting, adjusting for multiplicity amounts to putting equal prior mass on different sizes of models.  In contrast, the uniform prior for models involving $p$ covariates puts higher probability mass on larger models, reaching a maximum ${ p \choose 2 }$.  The details of the priors on the model space and parameters, including an empirical Bayes choice of the g-prior hyperparameter, are discussed in the appendix. 

Using this algorithm, we visit the most likely ETF factor models given our matrix of target assets.  Under the model and prior specification, there are closed-form expressions for the posteriors of the model parameters $\beta_{\gamma}$ and $\sigma$.  Thus, we can easily sample any functional of these parameters, including the implied tangency portfolio returns and Sharpe ratio.  Discussed in the empirical findings section, these metrics are useful in analyzing the selected ETF portfolios.  

\subsection{Derivation of the conditional loss function}

Our goal is now to describe the relationship between the ETFs and the unattainable assets. While the parameters of our model do precisely this, we have only posterior samples of these parameters (not a simple point estimate) and, moreover, these parameters are potentially ``larger" than we would like, in the sense that they involve all possible ETFs while perhaps a much smaller number accounts for the vast majority of the covariance structure of the target assets. 

In order to find a parsimonious summary we consider a loss function motivated by the conditional distribution of $\textbf{R}$ given $\textbf{X}$. In particular, this likelihood takes the form:

%
%

\begin{align}\label{like_loss}
	r \vert x \sim N(\gamma x,D^{-1}),	
\end{align}so the log-likelihood is:

\begin{align}
\log \det(D) - \frac{1}{2}\left(r^{T}Dr - 2x^{T}\gamma^{T}Dr + x^{T}\gamma^{T}D\gamma x\right).
\end{align}

Using this as our loss function, we might ask for an ``action" $\gamma$ that summarizes our distribution; we consider $D$ to be fixed. As we are not using this likelihood in a statistical capacity, we actually would like our $\gamma$ summary to characterize future realizations $\tilde{R}$ and $\tilde{X}$. Because these future realizations are naturally unavailable to use, we cannot maximize (\ref{like_loss}) over  $\tilde{R}$ and $\tilde{X}$. Instead, we first take expectations, yielding:
\begin{align}
\frac{1}{2} \text{tr}[D\Sigma_{r}] - \frac{1}{2}\text{tr}[\gamma^{T}D\gamma\Sigma_{x}] - \frac{1}{2}\mu_{x}^{T}\gamma^{T}D\gamma\mu_{x} + \text{tr}[\gamma^{T}D\beta\Sigma_{x}] + \mu_{x}^{T}\gamma^{T}D\mu_{r}.
\end{align}
Define the integrated conditional loss function, $\mathcal{L}(\gamma,\Sigma, \mu_{x},\mu_{y})$, by dropping all terms that do not involve our choice variable, $\gamma$:

\begin{align}
	\mathcal{L}(\gamma,\Sigma, \mu_{x},\mu_{r}) &= - \frac{1}{2}\text{tr}[\gamma^{T}D\gamma\Sigma_{x}] - \frac{1}{2}\mu_{x}^{T}\gamma^{T}D\gamma\mu_{x} + \text{tr}[\gamma^{T}D\beta\Sigma_{x}] + \mu_{x}^{T}\gamma^{T}D\mu_{r}.
\end{align}Of course, the parameters appearing in this expression are also not known exactly, so we integrate once more over the posterior distribution of$\lbrace\Sigma_{x},\beta, \mu_{x}, \mu_{y} \rbrace$:

\begin{align}
\mathcal{L}(\gamma) 
&= - \frac{1}{2}\text{tr}\left[D\gamma \left( \overline{\Sigma_{x}}  + \Sigma_{\mu_{x}} + \overline{\mu_{x}} \hspace{1mm} \overline{\mu_{x}}^{T} \right) \gamma^{T}\right] + \text{tr}\left[D \left( \overline{\beta\Sigma_{x}} + \Sigma_{\mu_{x}\mu_{r}} + \overline{\mu_{r}} \hspace{1mm} \overline{\mu_{x}}^{T} \right) \gamma^{T}\right]. \label{condlossfun}
\end{align}The overlines are used to represent the posterior means of the model parameters.  Defining $H = \overline{\Sigma_{x}} + \Sigma_{\mu_{x}} + \overline{\mu_{x}} \hspace{1mm} \overline{\mu_{x}}^{T} $, $f = \overline{\beta \Sigma_{x}} + \Sigma_{\mu_{x}\mu_{r}} + \overline{\mu_{r}} \hspace{1mm} \overline{\mu_{x}}^{T}$, and $H = LL^{T}$, we have:

\begin{equation}
\begin{split}
	\mathcal{L}(\gamma) &= - \frac{1}{2}\text{tr}\left(D\left[\gamma H\gamma^{T} - 2f\gamma^{T}\right]\right)
	\\
	&\propto - \frac{1}{2}\text{tr}\left(D\left[(\gamma-f H^{-1})H(\gamma-f H^{-1})^{T}\right]\right) \label{compsquare}
	\\
	&= - \frac{1}{2}\text{tr}\left((\tilde{\gamma}L-D^{\frac{1}{2}}f L^{-1})^{T}(\tilde{\gamma}L-D^{\frac{1}{2}}f L^{-1})\right)
	\\
	&= - \frac{1}{2}\text{\bf vec}\left(\tilde{\gamma}L-D^{\frac{1}{2}}f L^{-1}\right)^{T} \text{\bf vec}\left(\tilde{\gamma}L-D^{\frac{1}{2}}f L^{-1}\right).
\end{split}
\end{equation}
In \ref{compsquare}, we complete the square with respect to $\gamma$ and disregard constant terms that don't involve this action.  We also redefine the action as $\tilde{\gamma} = D^{\frac{1}{2}}\gamma$.  Finally, we convert the trace to an $l_{2}$ norm and distribute the vectorization operation across the expression using the Kroeneker product (where $\mathbb{I}$ is an identity matrix the same dimension as $D$):

\begin{equation}\label{lasso_form}
	\mathcal{L}(\tilde{\gamma}) = -\frac{1}{2} \norm{ \left[ [L^{T} \otimes \mathbb{I}]\text{\bf vec}(\tilde{\gamma}) - \text{\bf vec}(D^{\frac{1}{2}}f L^{-1}) \right] }_{2}^{2} + \lambda\norm{ \text{\bf vec}(\tilde{\gamma})}_{1}.
\end{equation}We include an $l_{1}$ penalty with parameter $\lambda$, which encourages the optimization solution to be sparse.  This is emphasized in Hahn and Carvalho's DSS paper where $l_{1}$-regularization is accompanied with integration over uncertainty for model selection.  


Expression \ref{lasso_form} is now in the form of standard sparse regression loss functions \citep{Tib}, with covariates $L$, ``data" $D^{\frac{1}{2}}f L^{-1}$, and regression coefficients $\tilde{\gamma}$. Accordingly we may optimize (\ref{lasso_form}) conveniently using existing software, such as the {\tt lars} package of \cite{Efron}.

Choice of the penalty parameter is a necessary practical concern.  In this matter, we also follow the pragmatic Bayesian approach of \cite{HahnCarvalho}, who advocate choosing $\lambda$ by scrutinizing plots that reflect the predictive deterioration attributable to $\lambda$-induced sparsification. Crucially, such plots convey posterior uncertainty in the chosen performance metric, allowing for intuitive criteria to be expressed along the lines of: chose $\lambda$ such that, with posterior probability greater than 95\%, the predictive error of the sparse predictor is no more than 10\% worse than that of the unsparsified optimal prediction." In our application, we will use the log conditional distribution as our measure of predictive performance, which is simply our utility function without the sparsity penalty.

\subsubsection{Difference from lasso and original DSS}

The loss function \ref{lasso_form} is distinct from original DSS loss function from \cite{HahnCarvalho} in two important ways. First, its derivation relies on a statistical model represented in the compositional form of conditional and marginal distributions, as opposed to a standard linear regression with normal i.i.d. errors.  Second, the loss metric is not explicitly squared error.  Instead, our notion of accuracy is defined by the negative log-likelihood of the conditional distribution of the target assets given the ETFs. Our approach is different from the group lasso of \cite{grouplasso} (where grouped covariates enter the model simultaneously along the lasso solution path) for these same reasons \cite{grouplasso}.  

In the appendix, \cite{HahnCarvalho} discusses covariance estimation in the context of a graphical lasso \citep{Friedmanetal2008} loss function, denoted as the ``DSS graphical model posterior summary optimization problem." The goal is to find a parsimonious posterior summary of the covariance using the graphical DSS loss function.  Our approach is similar, but our loss function is only focused on the off-diagonal block of the covariance quantifying the dependence between the target assets and the ETFs. That is, instead of considering the joint distribution, we focus on the implied conditional distribution. Unlike the graphical lasso optimization where all covariance components are penalized choice variables, our method allows for optimization only over coefficient matrix in the conditional distribution which represents the dependence between the ETFs and target assets.  This difference is made explicit by a simple example in the appendix.

An important feature of our loss function are the posterior means of cross products of parameters (such as $\overline{\beta\Sigma_{x}}$) that appear in our formulation.  They can have quite different distributions than the individual parameters as they are naturally dependent a posteriori via the model.  It is interesting and notable that these moments, quantifying the relationship between across such parameters, appear in our final loss function.

\section{Empirical Findings}



\begin{figure}[H]
\centering
  \includegraphics[scale=.4]{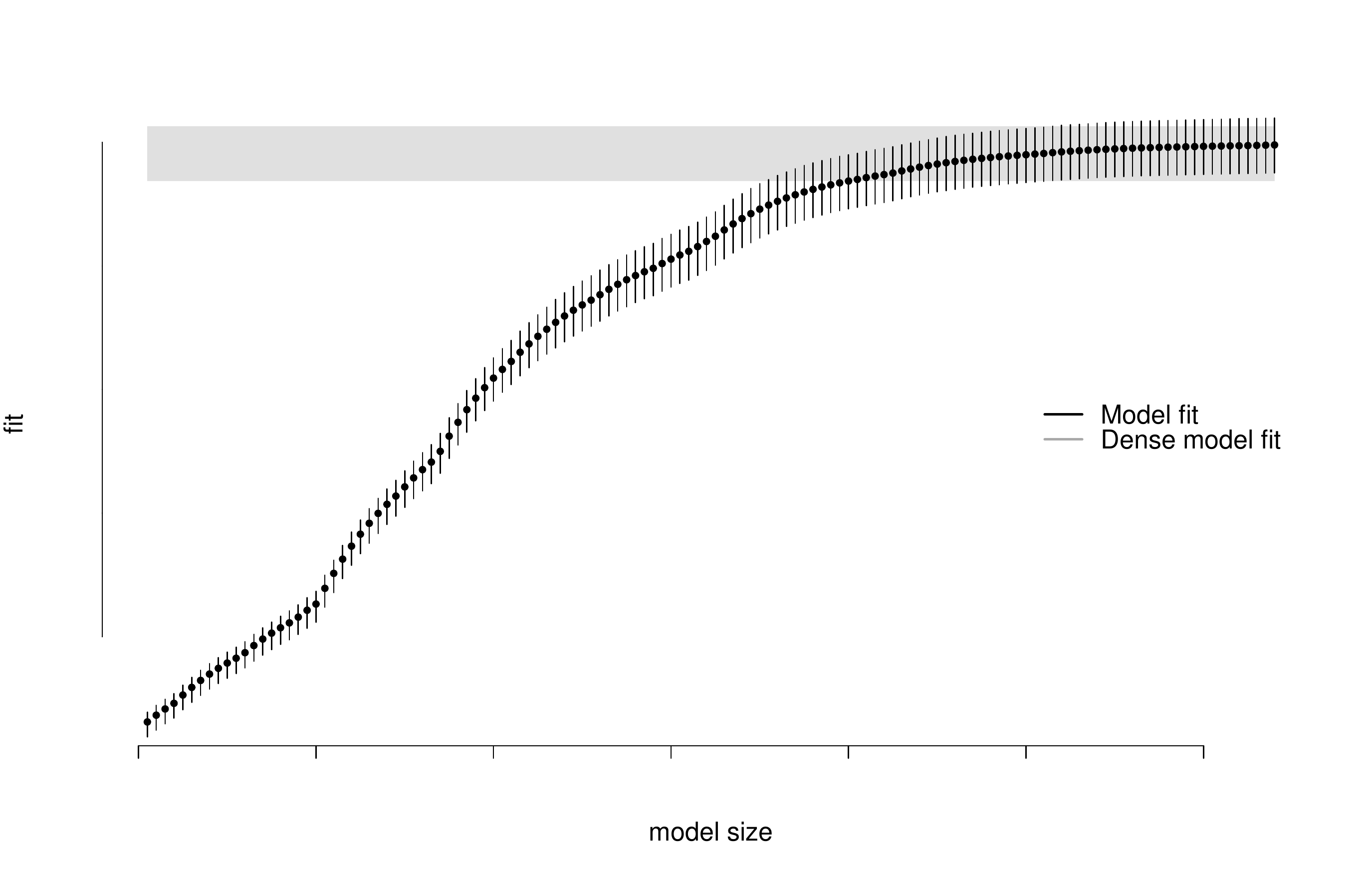}
  \caption{Model fits as measured by the conditional loss function. Allows for the selection of ETFs.  Model size refers to number of edges in graph.}
  \label{Lossgraph1}
\end{figure}

We now apply our model sampling and selection algorithm to ETF and financial anomaly data from February 1992 to February 2015. Using the parameters sampled in the matrix-variate MCMC, we calculate the value of our conditional loss function along the solution path of the lasso optimization and for several MCMC iterations at each solution. Recall that the loss function, written in ``lasso form," is:
\begin{equation}\label{lasso_form}
	\mathcal{L}(\tilde{\gamma}) = -\frac{1}{2} \norm{ \left[ [L^{T} \otimes \mathbb{I}]\text{\bf vec}(\tilde{\gamma}) - \text{\bf vec}(D^{\frac{1}{2}}f L^{-1}) \right] }_{2}^{2} + \lambda\norm{ \text{\bf vec}(\tilde{\gamma})}_{1}.
\end{equation} Figure \ref{Lossgraph1} shows the evaluations of this loss function for different values of $\tilde{\gamma}$ (and thus differing amounts of sparsity) and quantiles surrounding these evaluations. Note that model size is measured as the number of connections in the graph between the ETFs and target assets.  As model size gets larger, more ETFs connect to the target assets (more components of the conditional dependence matrix between the ETFs and target assets are nonzero) and the value of the conditional loss function, as derived by the log-likelihood, increases.  Every point along the ``model fit" line should be thought of as a graph representing the dependence between the ETFs and target assets.

The loss function value plateaus at the dense model fit, ie: when all possible edges between the ETFs and target assets sampled in our Gibbs algorithm are included.  The $40^{\text{th}}$ to $60^{\text{th}}$ quantile band of the dense model fit is shown in gray rectangle.  We remove the scale on the y-axis since we only need to compare the model fit relative to the dense model fit to make a graph selection.  Our heuristic for ETF selection is to choose the sparsest model on the solution such that the posterior mean of its fit is contained in the dense model quantile band.  This selection heuristic is key to our approach and can only be done if uncertainty intervals about the model fit our known.  Thus, a Bayesian fitting of our model provides these quantiles through which a selection can be made.  The model can be made sparser or denser (more or fewer edge connections between the ETFs and target assets) by varying the size of the quantile band.  This is a qualitative judgement to be made by the ``ETF selector."  However, since the first, model sampling, step tends to decrease the number of pertinent ETFs through exploration of only relevant models, we have found that a given ETF graph is relatively robust to changing the dense model fit quantiles.  

\begin{figure}[H]
\centering
  \includegraphics[scale=.55]{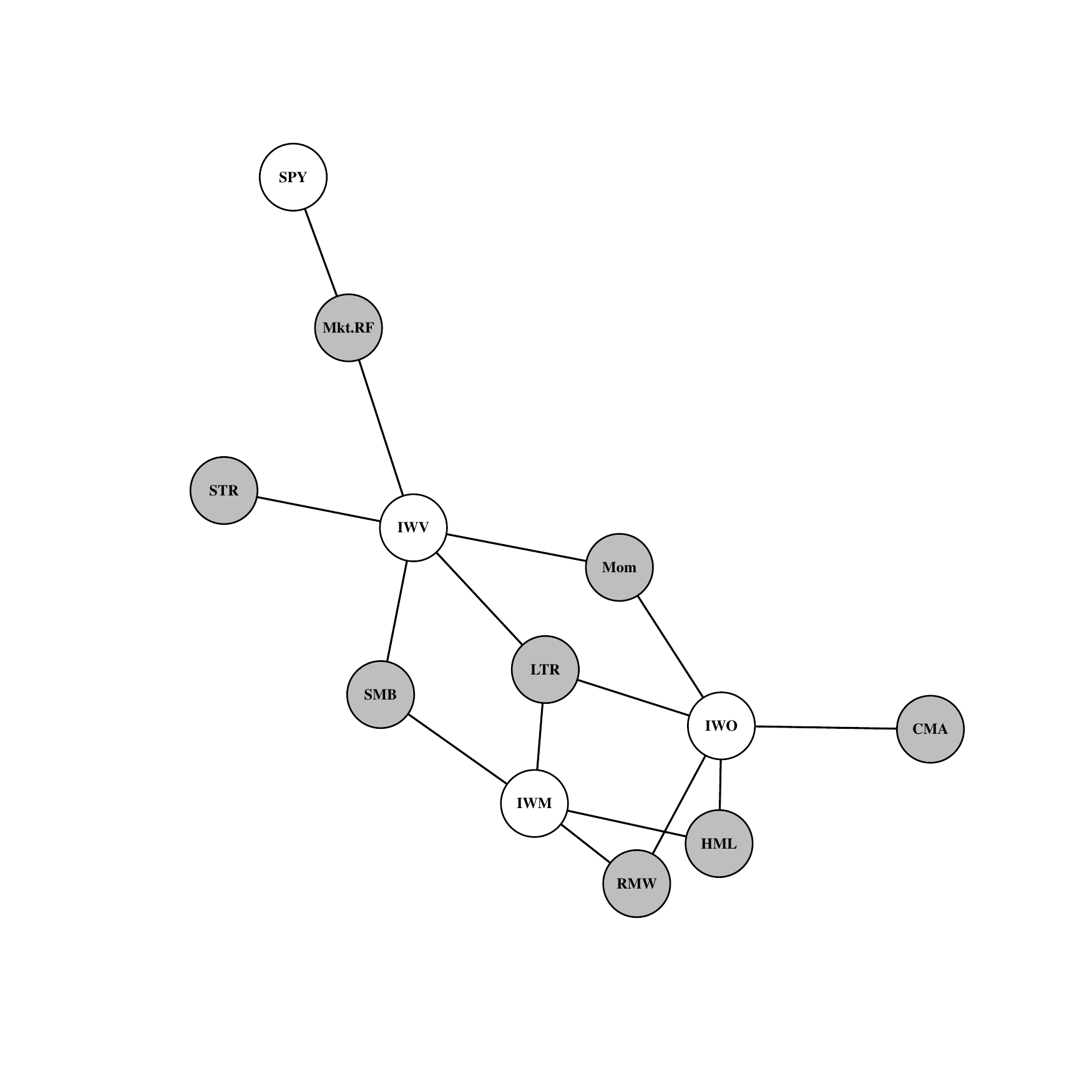}
  \caption{Selected ETFs and their edge connections to the unattainable assets.}
  \label{ETFgraph1}
\end{figure}

Figure \ref{ETFgraph1} shows the selected ETFs and their connection to the eight financial anomalies.  In the lasso optimization, we unpenalize the connection between SPY and the market factor (Mkt.RF) since an investor intuitively desires to hold at least ``the market."  SPY appears in all models along the solution path as a result of this unpenalization.  Note that IWM is connected to four of the eight anomalies.  Given that IWM is a small blend ETF, its connection to the SMB (small minus big, or size) factor is intuitive.  It is also connected to the LTR (long term reversal), HML (high minus low, or value), and RMW (robust minus weak, or profitability) factors.  Similar to STR (short-term reversal), LTR is a trading strategy that buys stocks that have had a below average long term trend of returns and sell stocks that have had an above average trend.  The intuition is that over time an outperforming (underperforming) stock will correct its above (below) average performance and systematically ``trend reverse." STR and LTR  capture the premium derived from this strategy over different length correction periods. IWM's connection to LTR suggests that small blend companies are exposed to the variation in LTR.  Additionally, its connection to HML and RMW indicates that some of the companies in IWM may be trading below their book value (ie: are ``value stocks") and that profitability is driving their return variation.  

IWO is an ETF comprised of small growth companies.  It is connected to the Mom (Momentum), CMA (conservative minus aggressive), LTR, RMW, and HML factors.  The Momentum factor invests in stocks that have had sustained outperformance that is expected persist.  IWO's connection to Mom is intuitive as the expansion of growing companies and their returns tend to be persistence through market cycles.  Its connection CMA is equally appealing.  CMA is a factor investing in companies who invest conservatively and selling companies who invest aggressively.  Small companies that are growing quickly are intimately involved in investment; whether it is increasing to fuel future growth or curbed to keep revenues high.  Therefore, IWO must be tied to a factor capturing market variation of companies focused on investment.  

The final ETF in our selected portfolio, IWV, is a blend of large market capitalization stocks.  Since this ETF contains companies and are large and relatively mature, it is a good substitute for a ``market-like" ETF.  Nonetheless, it is included in our selection along with SPY which tracks the S\&P500. As such, IWV and SPY are the only ETFs that are connected with the market factor.  Additionally, IWV is connected to STR, suggesting that large cap companies trend reverse or correct over short periods of time.  This makes sense when compared to IWM's connection to the LTR factor.  In general, larger companies such as Apple are more closely followed by the media and public compared to smaller companies.  Therefore, corrections to a large stock's above or below average performance should happen faster than a small stock.

\subsection{Benchmarking specific allocations}
In practice, individual investors want not only to know which funds to invest in, but also how much to invest in each.  Portfolio optimization and its vast literature is beyond the scope of this paper.  However, in this section we undertake an optimization approach that is Bayesian, intuitive, and simple.  We construct the selected ETF portfolio by maximizing the posterior mean of its Sharpe ratio.  The Sharpe ratio is a common financial metric characterizing the risk-adjusted return of an asset.  Acknowledging the widely used and reasonable assumptions that investors like high returns and dislike risk (are risk adverse), the Sharpe ratio divides the first two moments of an asset's return, i.e.: A higher Sharpe ratio indicates more return per unit of standard deviation (risk).  It was first mentioned in a paper written by nobel laureate William Sharpe in which he called it the ``reward-to-variability ratio" \citep{sharpe1966mutual}.

The space of weights is explored using a differential evolution optimization within the R package {\tt DEoptimR} of \cite{deoptimr}.  We are able to constrain the weights to sum to 100\% and restrict them to be positive (no short selling).  This is a reasonable constraint for a layman investor and one that we are able to easily enforce.  
The maximum posterior mean Sharpe ratio portfolio is 96\% market and large-cap ETFs with a 4\% tilt towards a small-cap ETF.  In sum, this portfolio includes ETFs that capture dominant sources of variation in the eight financial anomalies, and is allocated so that it risk-adjusted return is maximum.  It is reasonable to expect that market and large-cap ETFs will have large allocations in our portfolios.  These ETFs trade stocks that represent a large part of total market-cap of the U.S. financial markets.  The tilt towards small-cap and no allocation to growth suggests that we can increase our Sharpe ratio further with additional exposure to variation that drives the small minus big (SMB) factor.

\begin{table}
\begin{center}
\footnotesize
\begin{tabular}{|m{1cm}|c|c|c|c|}
\hline
\textbf{ETF} & SPY & IWM & IWO & IWV \\ \hline
\textbf{weight} & 59.3 \% & 4.0 \% & 0.0 \% & 36.7 \% \\ \hline
\textbf{style} & \textit{market} & \textit{small blend} & \textit{small growth} & \textit{large blend}  \\ \hline
\end{tabular}
\end{center}
\caption{February 1992 - February 2015: Selected ETF portfolio constructed by maximizing its Sharpe ratio's posterior mean (SPY is forced to be included through the lasso penalty).}
\label{ETFtable1}
\end{table}

\begin{figure}[H]
\centering
  \includegraphics[scale=.55]{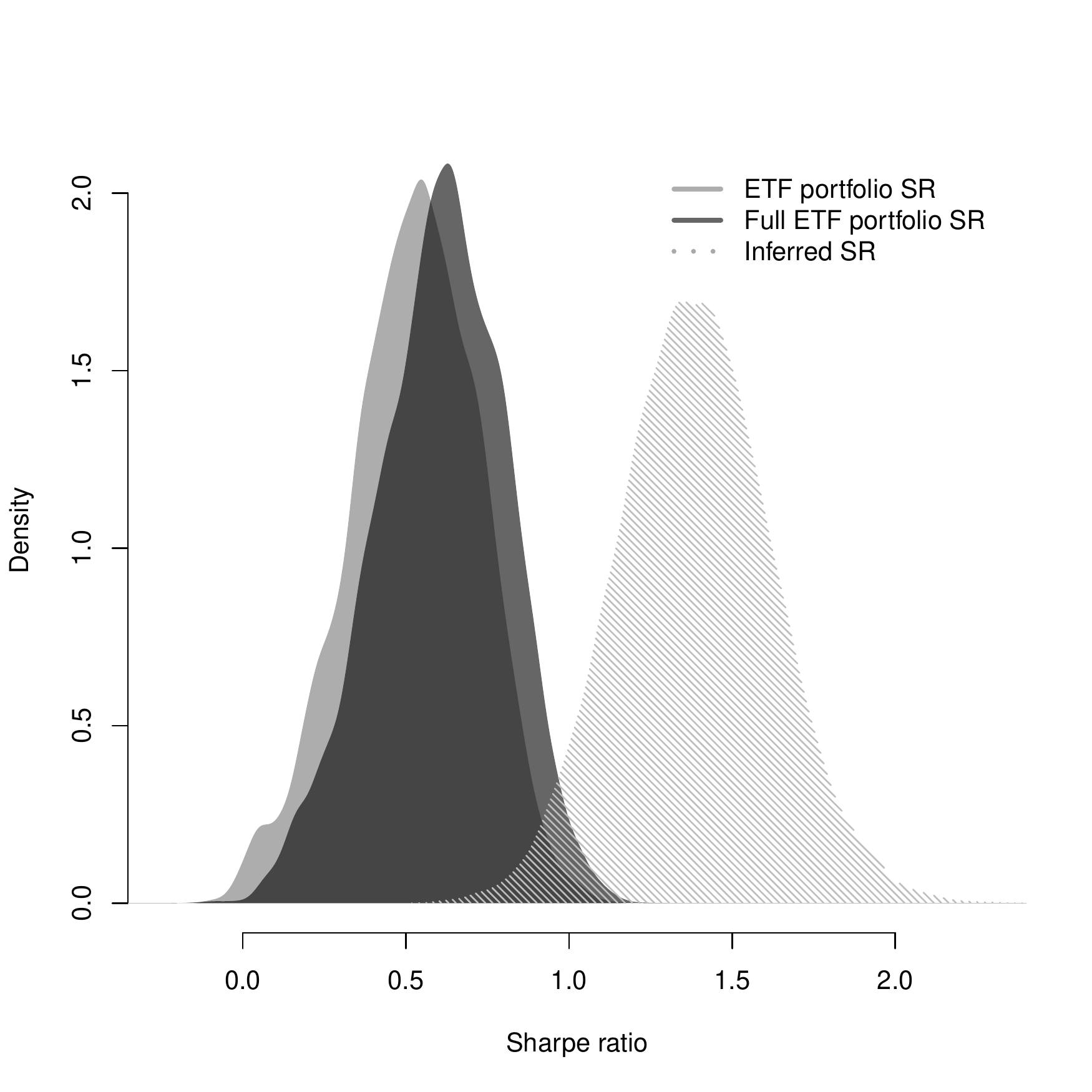}
  \caption{February 1992 - February 2015: Sampled Sharpe ratios for different portfolios}
  \label{ETFSR1}
\end{figure}

Crucially, what makes this orthodox Bayesian approach to allocation convenient and appealing was the initial reduction of the problem, first from all assets to only ETFs, and then, with our contribution, to just a small subset of ETFs. Therefore, it is natural to ask how much the variable reduced approach gives up to the analogous approach which forgoes the ETF selection step. We see this comparison in Figure \ref{ETFSR1}, which shows the distribution of sampled Sharpe ratios for the selected ETF portfolio and the full ETF porfolio (the maximum posterior mean Sharpe ratio portfolio of all 25 ETFs). We see that the distribution of Sharpe ratios for our reduced portfolio sits slightly below that of the distribution of Sharpe ratios of the portfolio investing in all ETFs (as one would expect). However, the two distributions overlap substantially and the price of an extremely parsimonious investing strategy is quite small.

\begin{figure}[H]
\centering
  \includegraphics[scale=.55]{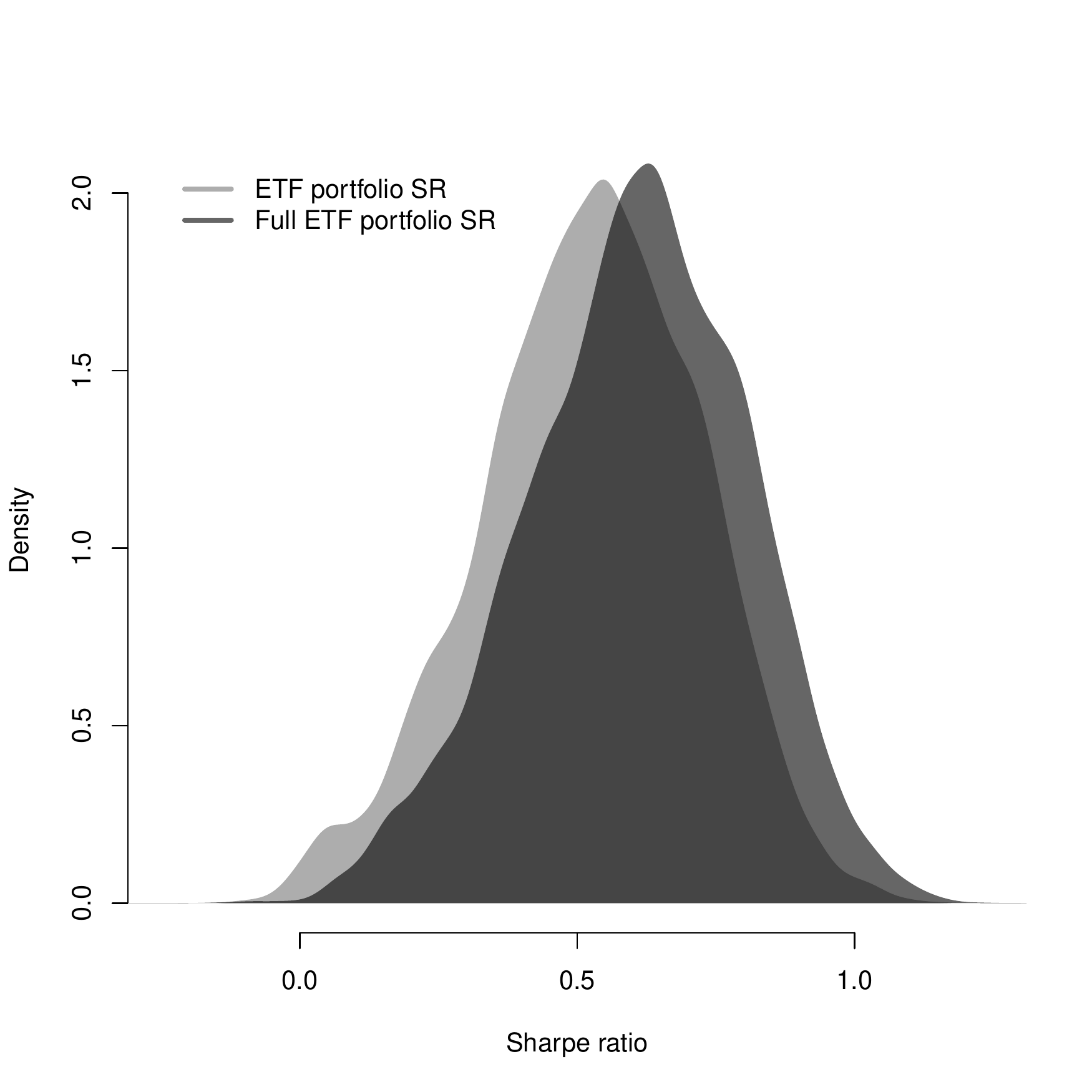}
  \caption{February 1992 - February 2015: Sampled Sharpe ratios for different portfolios}
  \label{ETFSR2}
\end{figure}

 Additionally, we consider the distribution of the Sharpe ratios corresponding to the model implied optimal portfolio if one could invest (both long and short positions) in the target assets themselves.  This benchmark is unattainable in two ways.  First, our analysis was predicated on the idea that the left-hand-side assets could not be directly invested in. More importantly, the Sharpe ratios obtained are from an optimal model that is allowed to change iteration-by-iteration during posterior sampling.  That is, we are looking at the distribution of the performance of the optimal portfolio and not the performance of the optimal portfolio in expectation; it is not a single portfolio we are looking at, but many conditionally optimal ones. So, while the inferred SR optimal portfolio cannot be achieved, it does provide a natural scale for our comparisons.   In figure \ref{ETFSR2}, we show the same sampled Sharpe ratios with the inferred distribution removed.

\begin{figure}[H]
\centering
  \includegraphics[scale=.55]{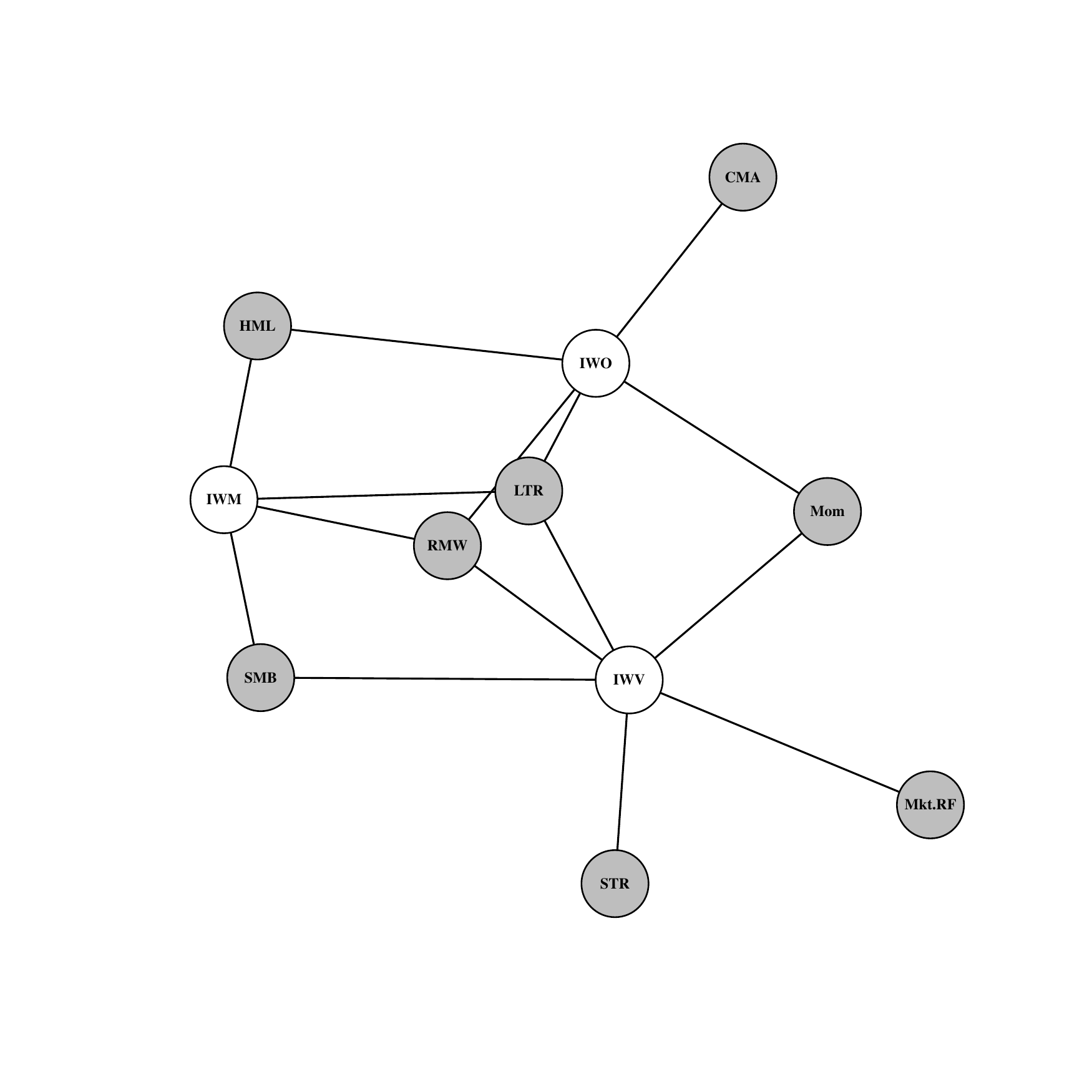}
  \caption{Selected ETFs and their edge connections to the unattainable assets.}
  \label{ETFgraph2}
\end{figure}

Lastly, we consider the case where the market ETF (SPY) is treated equal to any other ETF, and so can be dropped from the select set.  The resulting ETF graph is shown in figure \ref{ETFgraph2}.  This graph is the same as figure \ref{ETFgraph1} except that SPY falls out due to penalization. The maximum posterior mean Sharpe ratio portfolio the contains only one ETF, IWV.  This is a broad market ETF tracking the Russell 3000 index.  After integrating over uncertainty in future returns and parameters, the chosen ETF portfolio is simply a market-like ETF.  This provides rigorous intuition for the folk wisdom ``just buy the market."  After taking all unknowns into account, this result suggests that a broad portfolio of large stocks is not too bad. 

\begin{table}[H]
\begin{center}
\footnotesize
\begin{tabular}{|m{1cm}|c|}
\hline
\textbf{ETF} & IWV \\ \hline
\textbf{weight} & 100 \% \\ \hline
\textbf{style} & \textit{large blend}  \\ \hline
\end{tabular}
\end{center}
\caption{February 1992 - February 2015: Selected ETF portfolio constructed by maximizing its Sharpe ratio's posterior mean (SPY is not forced to be included through the lasso penalty).}
\label{ETFtab1}
\end{table}

Although confirming one widely help position, this result raises the question as to why there are other camps who espouse alternative investing advice. In our next section, we see that non-stationarity can explain some of this discrepancy.

\subsection{Rolling analysis}

\begin{figure}[H]
\centering
\subfigure[March 1995 - February 2005]{\includegraphics[width=3.23in]{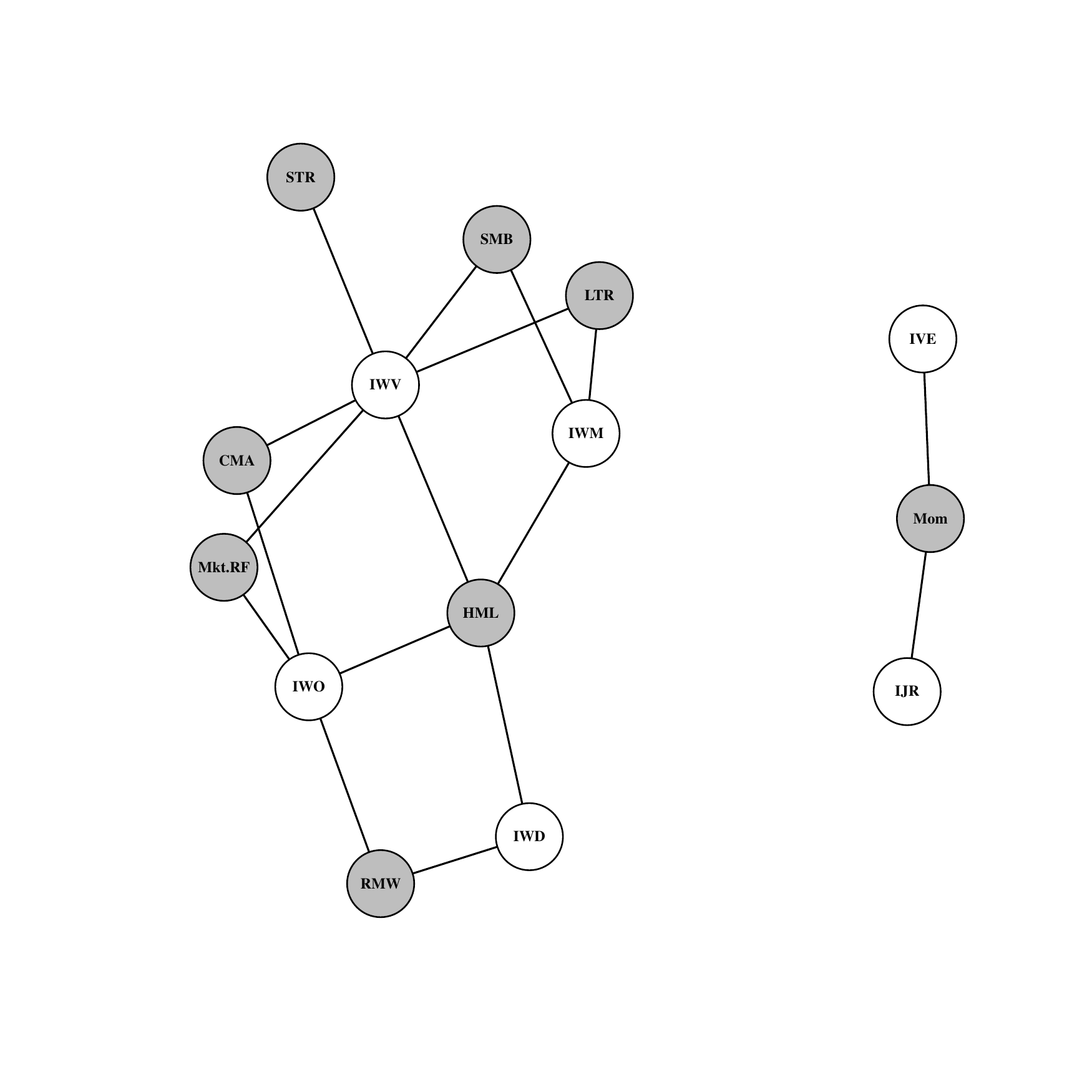}	}  
\subfigure[March 2000 - February 2010]{\includegraphics[width=3.23in]{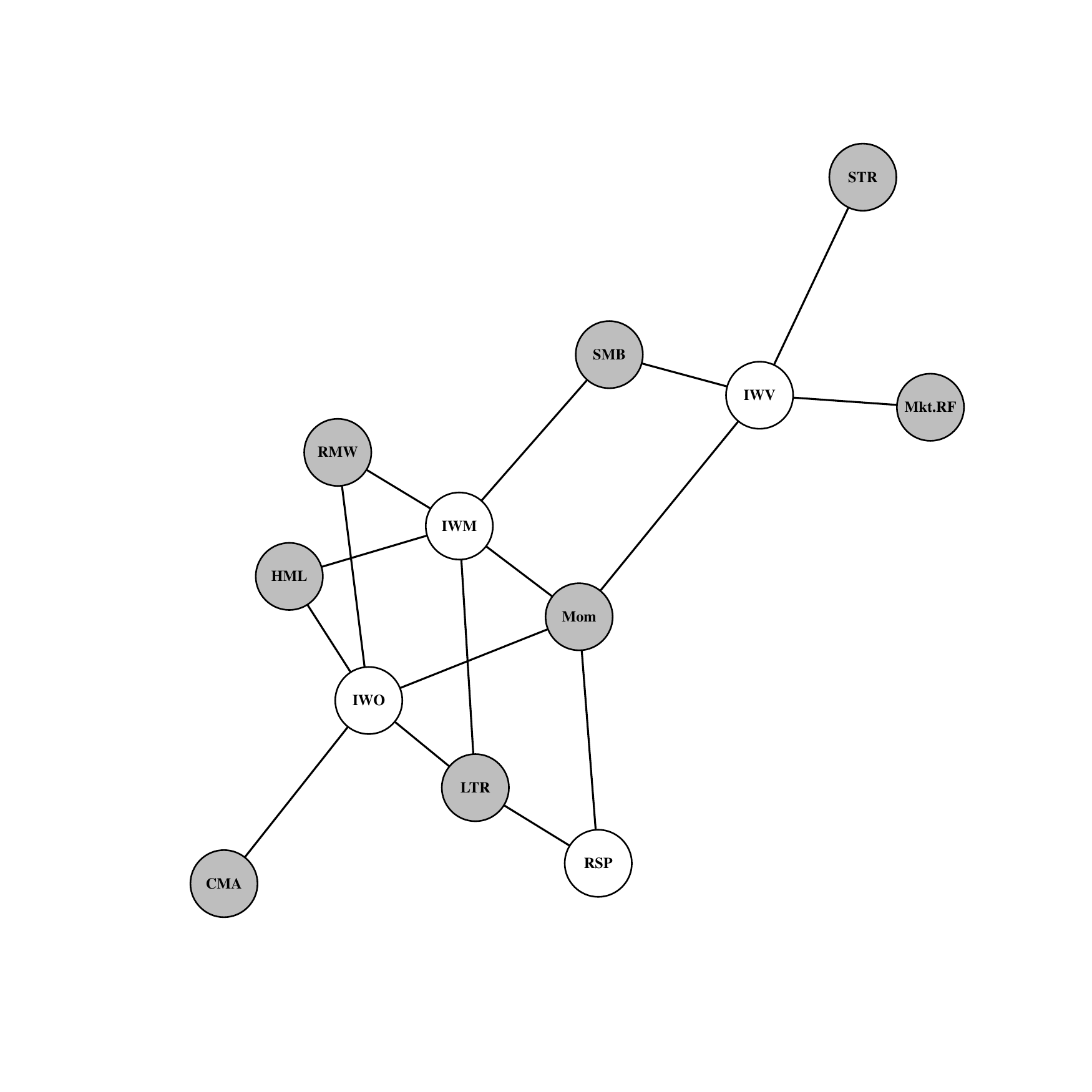}}
\subfigure[March 2005 - February 2015]{\includegraphics[width=3.23in]{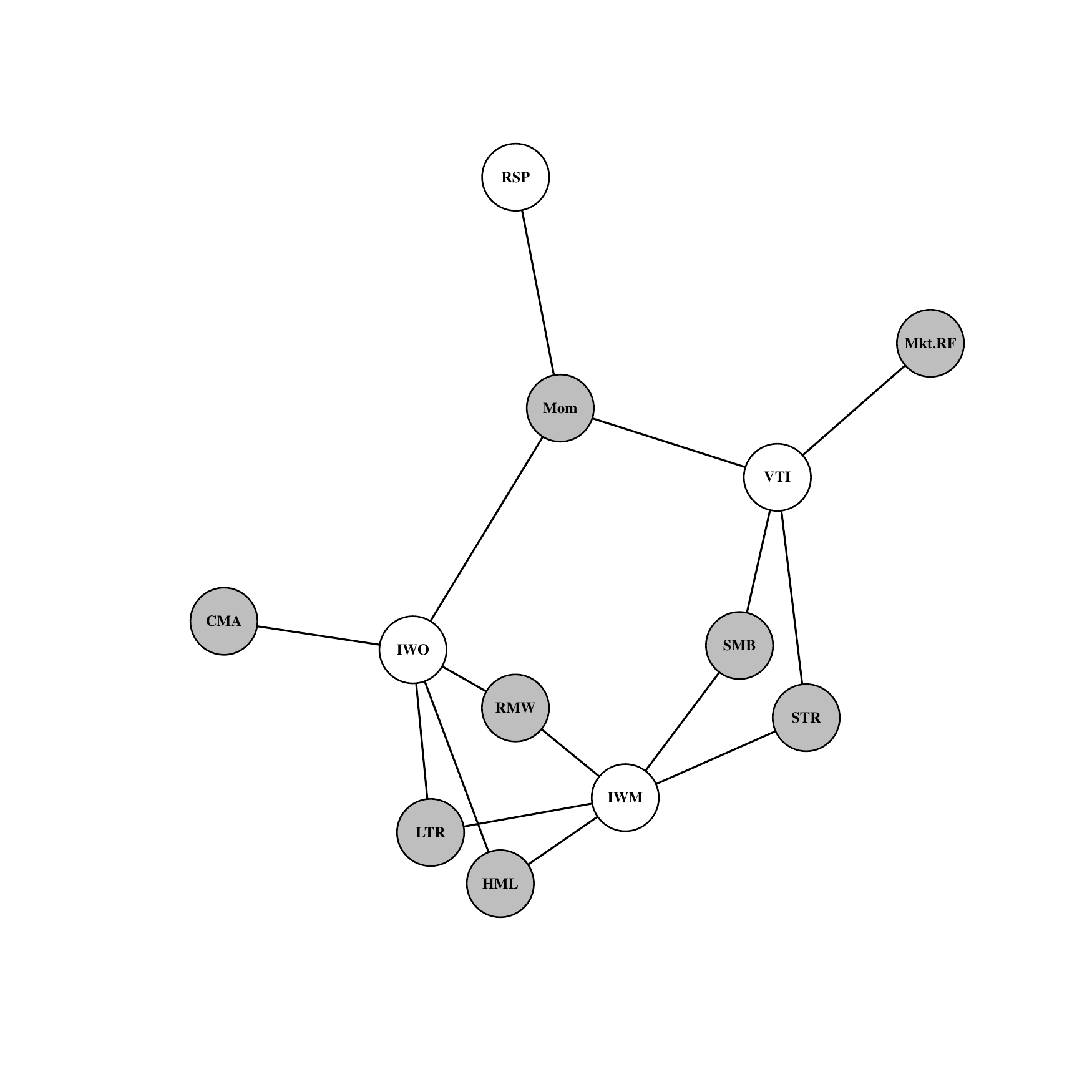}}
\label{rollgraphs}
\end{figure}

Our analysis so far, built on an APT model, has assumed stationarity of the returns process, meaning that the distribution of returns are stable across time. It is natural to question this assumption. To investigate the possibility that the returns vary in time, and to examine how this might impact our ETF selection method, in this section we apply our approach separately to overlapping time periods..  We show overlapping 10-year periods from March 1995 through February 2015 in figure \ref{rollgraphs}.  More ETFs become available in the different time periods and we allow the algorithm to consider the larger set.  The first time period has 25 ETFs and the final time period has 46 ETFs.  Each ETF set is a subset of the future time periods' data.  For the first  time period (March 1995 - February 2005), we see value appear in IWD and IVE as well as small-cap in IJR.  Otherwise, the usual large-stock blend appears in IWV and small-stock blends in IWO and IWM.  Notice how momentum is detached from the broad market ETFs.  This suggests that momentum-based trading had isolated covariation among only a couple ETFs.  Just before the beginning of this time period, \cite{jegadeesh1993returns} published their famous paper on the momentum strategy and a formalization of those ideas were in their infancy.  Also, note that the small growth ETF, IWO, is connected to the market factor during this time period.  March 1995 to February 2005 includes the tech boom and burst where many small technology companies grew at enormous rates.  This ultimate bubble caused significant variation in the markets, and we see this manifested through IWO's connection to the market factor.    

The two more recent time periods are similar.  Both have three in common (IWO, IWM, RSP) and each has a broad market ETF (IWV for March 2000 - February 2010 and VTI for March 2005 - February 2015). The factors' connections to the ETFs do change through the periods, highlighting the reasonable fact that covariation within a given ETF may be driven by different factors over time.  Specifically, we see that IWO loses its connection to the market factor and IWM gains connections to RMW and Mom.  Also, the momentum factor joins the larger graph with connections to all selected ETFs suggesting that its variation is matured and exists in the broader market.  The S\&P500 equal weight ETF, RSP, enters through the momentum and long-term reversal factors during the second time period. 

One interesting comparison is the March 2005 - February 2015 portfolio (table \ref{ETFtab3}) and the portfolio formed over the longer time period, February 1992 - February 2015, in table \ref{ETFtab1}.  The former results in a portfolio of a single ETF of a blend of large stocks. In the latter, there is a roughly even split between broad market and market equal-weight ETFs.  This latter portfolio is tilted toward smaller stocks giving the equal weighting of RSP and is due to the shorter and more recent data used. 

\begin{table}[H]
\begin{center}
\footnotesize
\begin{tabular}{|m{1cm}|c|c|c|c|}
\hline
\textbf{ETF} & VTI & IWM & IWO & RSP \\ \hline
\textbf{weight} & 43.7 \% & 0.0 \% & 0.0 \% & 56.3 \% \\ \hline
\textbf{style} & \textit{broad market} & \textit{small blend} & \textit{small growth} & \textit{market equal-weight}  \\ \hline
\end{tabular}
\end{center}
\caption{March 2005 - February 2015: Selected ETF portfolio constructed by maximizing its Sharpe ratio's posterior mean}
\label{ETFtab3}
\end{table}

\subsection{Further Applications}

\subsubsection{Comparison to Wealthfront}

\begin{figure}[H]
  \includegraphics[width=0.8\linewidth]{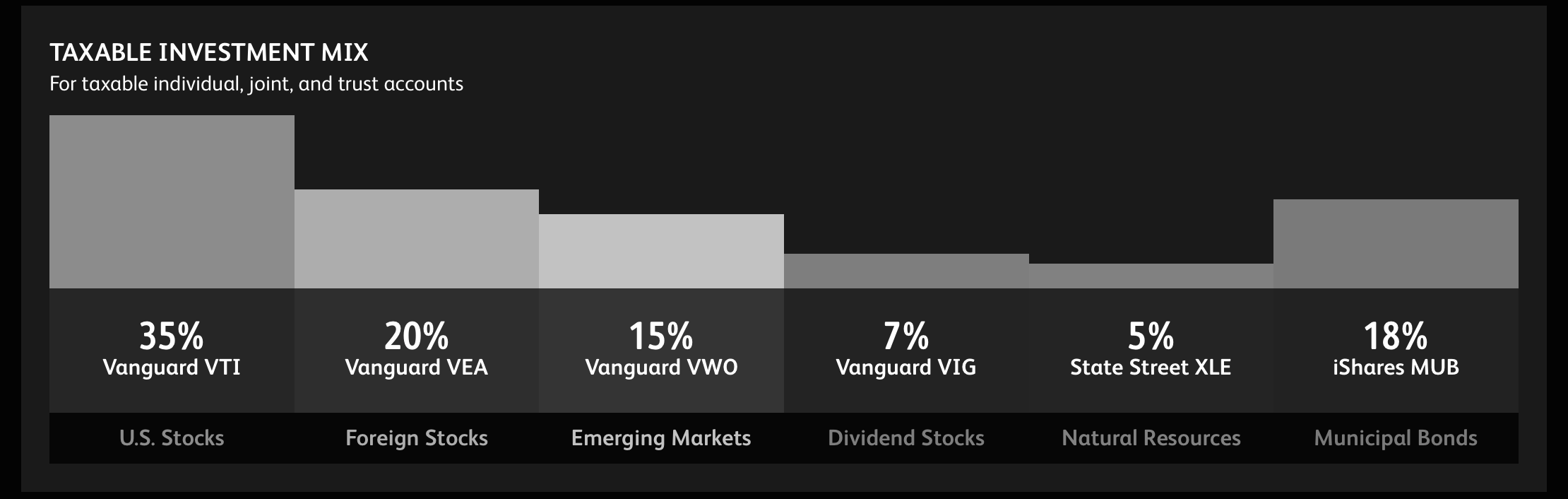}
  \caption{Source: wealthfront.com}
  \label{wealth}
\end{figure}

\begin{figure}[H]
\centering
  \includegraphics[scale=.5]{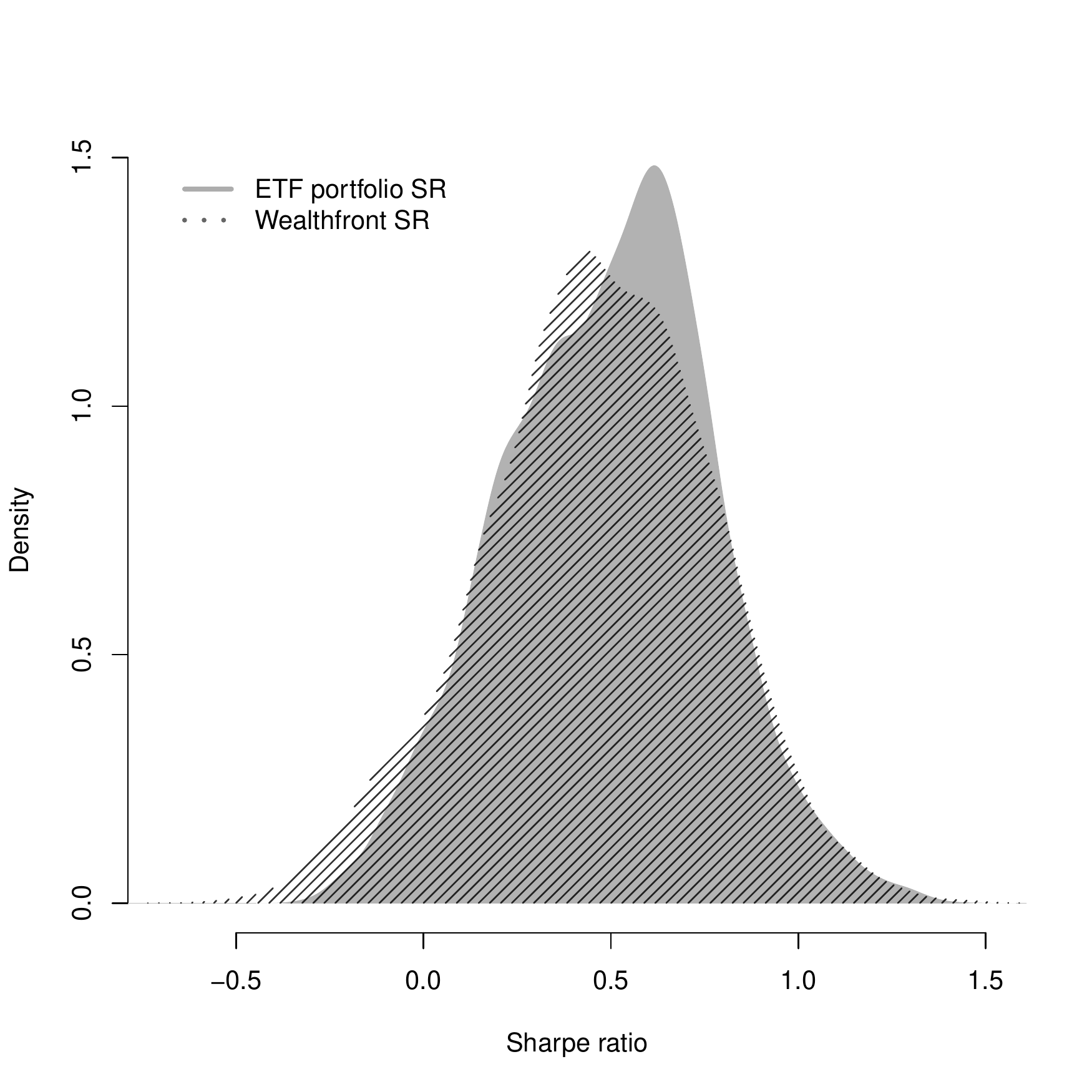}
  \caption{March 2005 - February 2015: Sampled Sharpe ratios for Wealthfront portfolio and our proposed ETF portfolio given in table \ref{ETFtab3}.}
  \label{SRWF}
\end{figure}

Over the shorter data period (March 2005 - February 2015), we are able to compare our selected portfolio to Wealthfront.com, an ETF investing firm that currently has \$2.6bn in assets under management.   In the past five years as ETFs have become increasingly popular, many other firms and products similar to Wealthfront's have emerged, including: Betterment, Charles Schwab Intelligent Portfolios, WiseBanyan, and LearnVest.  Each company is marketed to the average investor.  Given a level of risk tolerance determined by a series of questions answered by the investor, each company will generate a corresponding portfolio of ETFs. Figures \ref{wealth} shows an example allocation from Wealthfront's website.  Note the presence of only ETFs in each portfolio; these are portfolios of this one passive instrument.  The growth of these companies is driven by the desire of the layman investor to not just index invest, but to index invest in an optimal way.  Even though these companies make a reasonable first pass at providing a solution - to our knowledge, their product falls short in two key areas. (1) Their selection of ETFs is based on a qualitative analysis of expense ratios, liquidity, general market popularity, and predefined asset class buckets.  (2) The optimal weights are calculated from traditional and potentially unstable mean-variance optimization.  In this paper, we attempt to provide an improvement to step (1).   
We display the weights of the equity-only portfolio given by Wealthfront in table \ref{WFPort}.  In figure \ref{SRWF}, we display the sampled Sharpe ratios.  Both portfolios have similar upside potential, but note that the Wealthfront portfolios left tail is substantially larger than the ETF portfolio. 
\begin{table}
\begin{center}
\footnotesize
\begin{tabular}{|m{1cm}|c|c|c|c|c|}
\hline
\textbf{ETF} & VWO & VEA & VTI  & VIG & XLE \\ \hline
\textbf{weight} & 18.3 \% & 24.4 \% & 42.7 \% & 8.6 \% & 6 \% \\ \hline
\textbf{style} & \textit{EM} & \textit{Non-US} & \textit{market} & \textit{dividend} & \textit{energy} \\ \hline
\end{tabular}
\end{center}
\caption{Wealthfront portfolio.}
\label{WFPort}
\end{table}

\subsubsection{Mutual funds as target assets}

\begin{figure}[H]
\centering
  \includegraphics[scale=.5]{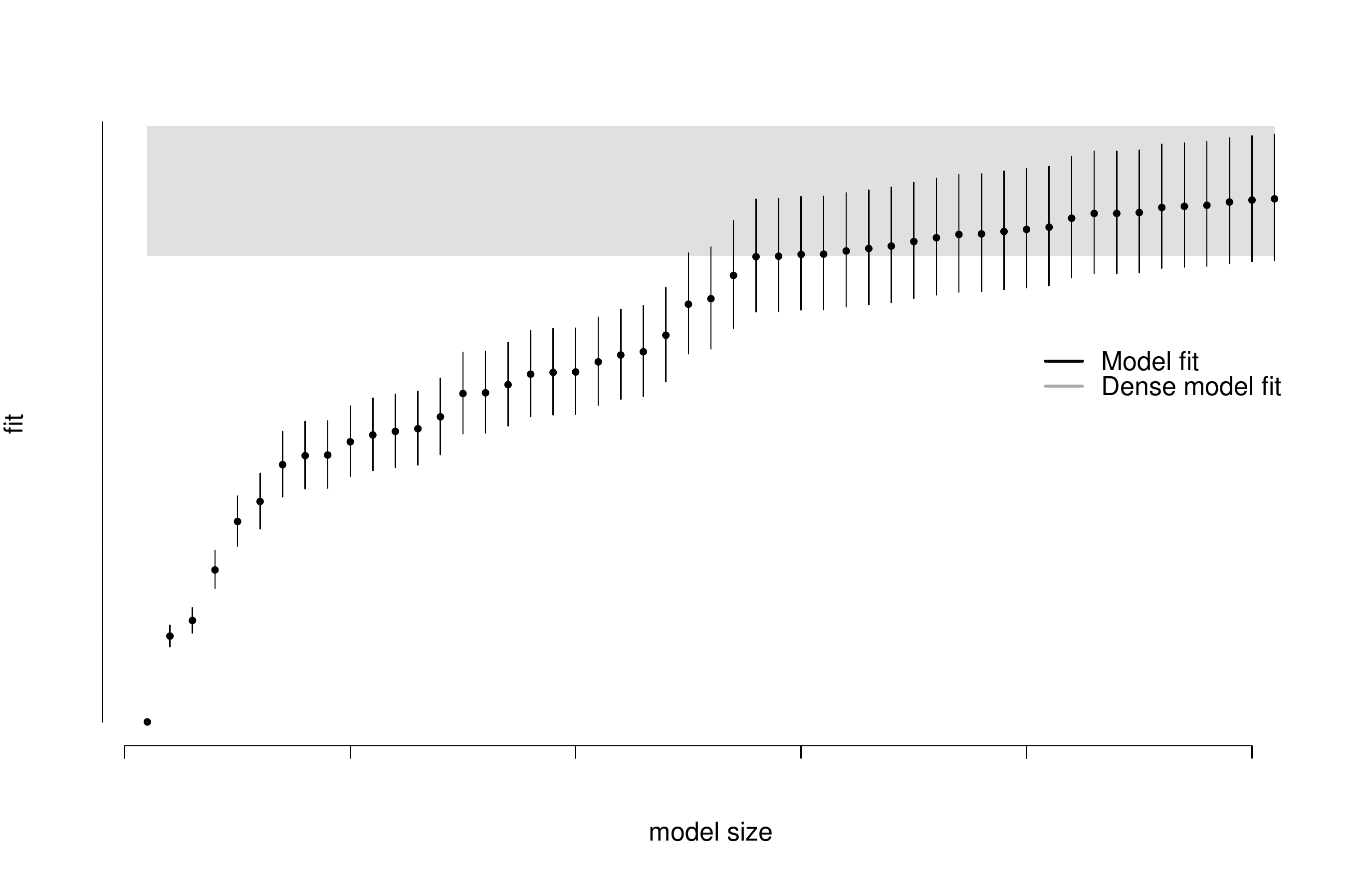}
  \caption{Model fits as measured by the conditional loss function. Allows for the selection of ETFs.  Model size refers to number of edges in graph.}
  \label{Lossgraphmf}
\end{figure}

\begin{figure}[H]
\centering
  \includegraphics[scale=.75]{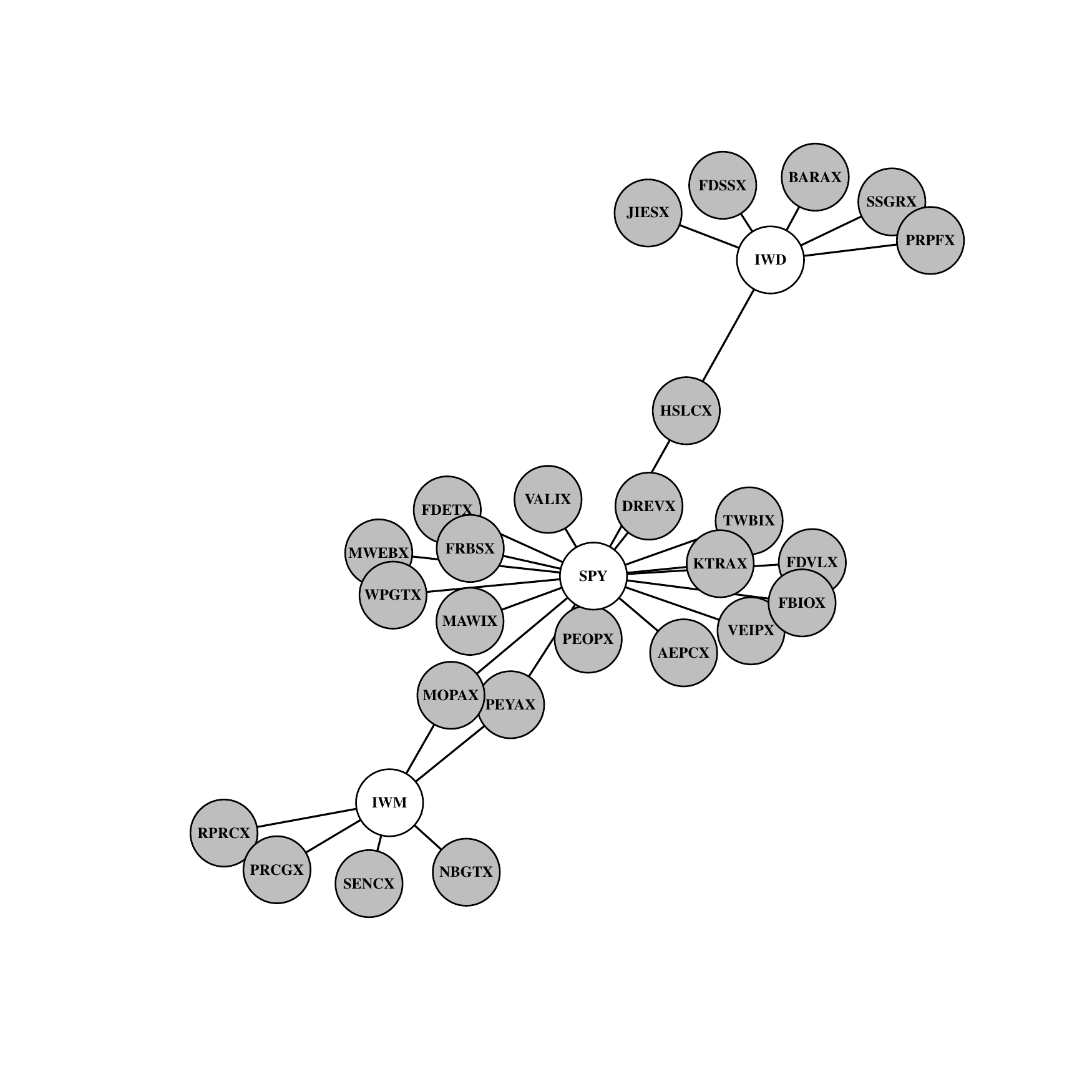}
  \caption{Selected ETFs and their edge connections to the set of mutual funds.  Singleton mutual funds (with no edges) are not shown for clarity.}
  \label{ETFMFgraph}
\end{figure}

As a final exercise, we consider the case of having mutual funds as target assets.  We randomly sample 100 mutual funds from the CRSP Survivor-Bias-Free US Mutual Fund database and use these as our response matrix in our algorithm and our data is over the longer time period, February 1992 - February 2015. The solution path allowing us to select the appropriate model is shown in figure \ref{Lossgraphmf}, and the selected graph is displayed in figure \ref{ETFMFgraph}.  
The selected graph shows the mutual funds that are connected to the chosen ETFs, and the result is quite remarkable.  The three chosen ETFs have strategies precisely linked to the Fama and French three factors from their well known paper \cite{FF3}. This suggests that the covariation among mutual fund returns is largely encompassed by variation in the market, size, and value factors. However, the direction of causation is unknown.  Either these three factors represent the true dimensions of the financial market, or mutual fund managers believe these are the dimensions and trade as such.  

This example emphasizes the broader value and applicability of our algorithm.  In today's world, there are thousands of mutual funds and ETFs one could invest in, and an investor is quickly overwhelmed with bank research, Morningstar ratings, and qualitative advice on which small subset of investments she should care about.  Using modern technology in Bayesian estimation and regularized optimization combined with economic theory in the APT, our selection algorithm is able to sparsify this massive set of investment options for the average investor.

\newpage
\section{Conclusion}

The investment universe is complicated.  With the rise of multi-billion dollar money managers, active and passive mutual funds spanning every conceivable asset class, and several varieties of financial products and hedging instruments, it is challenging for the average investor to find the best place to invest.  A common answer to this investment question is: "Just hold the market."  

However, this simple solution might not be so simple to implement.  How does one define the market?  Is it all public equity?  If so, it is impossible to hold each of these assets.  The Standard and Poor's 500 index (S\&P 500) comprised of the top 500 largest companies traded on the major exchanges might be a decent proxy for the United States equity market, but does it capture all of the market variation?  Further, is there a way to identify which premia derived from this variation are most important?  Answers to these complicated questions would greatly benefit the average investor.

An investment that gives broad exposure to many asset classes for low fees is the exchange traded fund (ETF).  Are ETFs a working solution to the "just hold the market" dilemma?  Perhaps.  As the ETF universe expands into new financial markets and asset classes, there is no doubt that the average investor is exposed to more market variation than ever before.  In fact, ETFs have introduced a secondary but important wrinkle in the investing decision: in which ETFs do I invest, and how much should I invest in each?  The investment problem becomes a choice of ETFs and portfolio allocation problem.  This index investing dilemma has led to the development of two notable firms in the past five years \footnote{Others include Charles Schwab Intelligent Portfolios, WiseBanyan, and LearnVest.}: Betterment and Wealthfront.  Each company is marketed to the average investor.  Given a level of risk tolerance determined by a series of questions answered by the investor, each company will generate a corresponding portfolio of ETFs.    
%
%

In this paper, we proposed a methodology to address these shortfalls of index investing.  We formulate, from the investor perspective, the ETF choice problem as a Bayesian model selection problem where we select ETFs that most closely replicate a chosen set of target assets.  We lean on the theoretical underpinnings of stochastic search variable selection from \cite{GeorgeandMcCulloch} and arbitrage pricing theory of \cite{Ross} to develop our method. We then couple our statistical analysis with a practical variable selection approach based on decision theory, the end result being a handful of ETFs from which to build a portfolio.  Crucially, our analysis does not stop there. We may continue to leverage the insights of our fully Bayesian statistical analysis to benchmark various portfolio allocations (among the selected ETFs) on the basis of widely-used criteria, such as the Sharpe ratio.  

An important point to remember when considering asset allocation is that not only are future returns uncertain, but the distributions of those future returns are likewise uncertain.  Additionally, while we might want to compare our ETF portfolio to the optimal portfolio, this optimal portfolio is itself unknown (in addition to being impracticable).  Fortunately, our Bayesian analysis permits us to compare the performance of any candidate portfolio to the unknown optimal portfolio, while accounting for all of these many sources of uncertainty. We make these comparisons manageable by first undertaking a principled variable selection step.

The upshot of our analysis is both expected and surprising.  On the one hand, we find that, up to statistical uncertainty, our chosen ETF portfolios contain only a couple of broad-spectrum ``market" ETFs.  That is, as far as our data inform us, the most sensible ETF portfolios to hold are largely exposed to the market.  Moreover, the broad market index the algorithm chooses is not SPY, but IWV: an ETF composed of large-cap stocks.  We also find that our selected ETF portfolios have similar Sharpe ratio profiles to an unreasonable alternative of investing in all available ETFs.  Over rolling time periods, our analysis routinely chooses a more diverse set of ETFs indicating that one should adjust their portfolios as markets change.  Indeed, our analysis tilts heavily towards small-cap and value funds, with a dash of equal-weighting through RSP tied to the momentum factor. This finding lends statistical credence to prevailing folk-wisdom in investing circles.

\newpage
\appendix

\section{Matrix-variate Stochastic Search}

For model comparison, we calculate the Bayes factor with respect to the null model without any covariates.  First, we calculate a marginal likelihood.  This likelihood is obtained by integratingthe full model over $\boldsymbol \beta_{\gamma}$ and $\sigma$ multiplied by a prior for these parameters.  A Bayes factor of a given model $\gamma$ versus the null model, $B_{\gamma 0} = \frac{m_{\gamma}\left(\textbf{R}\right)}{m_{0}\left(\textbf{R}\right)}$ with:

\begin{align} \label{marginal}
m_{\gamma}\left(\textbf{R}\right) = \int \textrm{MN}_{T,q}\left( \textbf{R} \hspace{1mm} \vert \hspace{1mm} \textbf{X}_{\gamma} \boldsymbol \beta_{\gamma}, \hspace{1mm} \sigma^{2}\mathbb{I}_{\textrm{T x T}}, \hspace{1mm} \mathbb{I}_{\textrm{q x q}}\right) \pi_{\gamma}\left(\boldsymbol \beta_{\gamma}, \sigma\right) d\boldsymbol \beta_{\gamma} d\sigma.
\end{align}From the APT assumption, we have that the columns of $\textbf{R}$ are independent.  Additionally, we assume independence of the priors across columns of $\textbf{R}$ so we can write the integrand in \ref{marginal} as a product across each individual target asset:

\begin{align*}
m_{\gamma}\left(\textbf{R}\right) &= \int \Pi_{i=1}^{q} \hspace{1mm} \textrm{N}_{T}\left( \textbf{R}^{i} \hspace{1mm} \vert \hspace{1mm} \textbf{X}_{\gamma} \boldsymbol \beta_{\gamma}^{i}, \hspace{1mm} \sigma^{2}\mathbb{I}_{\textrm{T x T}}\right) \pi_{\gamma}^{i} \left(\boldsymbol \beta_{\gamma}^{i}, \sigma\right) d\boldsymbol \beta_{\gamma}^{i} d\sigma
\\
&\iff
\\
m_{\gamma}\left(\textbf{R}\right) &= \int \hspace{1mm} \textrm{N}_{T}\left( \textbf{R}^{1} \hspace{1mm} \vert \hspace{1mm} \textbf{X}_{\gamma} \boldsymbol \beta_{\gamma}^{1}, \hspace{1mm} \sigma^{2}\mathbb{I}_{\textrm{T x T}}\right) \pi_{\gamma}^{1} \left(\boldsymbol \beta_{\gamma}^{1}, \sigma\right) d\boldsymbol \beta_{\gamma}^{1} d\sigma \\ & \times \cdots \times \int \textrm{N}_{T}\left( \textbf{R}^{q} \hspace{1mm} \vert \hspace{1mm} \textbf{X}_{\gamma} \boldsymbol \beta_{\gamma}^{q}, \hspace{1mm} \sigma^{2}\mathbb{I}_{\textrm{T x T}}\right) \pi_{\gamma}^{q} \left(\boldsymbol \beta_{\gamma}^{q}, \sigma\right) d\boldsymbol \beta_{\gamma}^{q} d\sigma
\\
&= m_{\gamma}\left(\textbf{R}^{1}\right) \times \cdots \times m_{\gamma}\left(\textbf{R}^{q}\right) 
\\
&= \Pi_{i=1}^{q} m_{\gamma}\left(\textbf{R}^{i}\right),
\end{align*}

with:

\begin{align} \label{A5}
\textbf{R}^{i} \sim \textrm{N}_{T}\left(\textbf{X}_{\gamma} \boldsymbol \beta_{\gamma}^{i}, \hspace{1mm} \sigma^{2}\mathbb{I}_{\textrm{T x T}}\right).
\end{align}Therefore, the Bayes factor for this matrix-variate model is just a product of Bayes factors for the individual multivariate normal models - a direct result of the APT model assumptions.

\begin{align} \label{A6}
B_{\gamma0} = \widetilde{B}_{\gamma0}^{1} \times \cdots \times  \widetilde{B}_{\gamma0}^{q}
\end{align}

with:

\begin{align} \label{A6}
\widetilde{B}_{\gamma0}^{i} = \frac{m_{\gamma}\left(\textbf{R}^{i}\right)}{m_{0}\left(\textbf{R}^{i}\right)}.
\end{align}

The simplification of the marginal likelihood calculation is crucial for analytical simplicity and for the resulting SSVS algorithm to rely on techniques already developed for vector response models.  In order to calculate the integral for each Bayes factor, we need priors on the parameters $\beta_{\gamma}$ and $\sigma$.  Since the priors are independent across the columns of $\textbf{R}$, we aim to define $\pi_{\gamma}^{i} \left(\boldsymbol \beta_{\gamma}^{i}, \sigma\right)$ $\forall i \in \{1,...,q\}$, which we express as the product: $\pi_{\gamma}^{i} \left(\sigma\right) \pi_{\gamma}^{i} \left(\boldsymbol \beta_{\gamma}^{i} \hspace{1mm} \vert \hspace{1mm} \sigma\right)$.  Motivated by the work on regression problems of Zellner, Jeffreys, and Siow, we choose a non-informative prior for $\sigma$ and the popular g-prior for the conditional prior on $\boldsymbol \beta_{\gamma}^{i}$, \citep{Z1}, \citep{Z2}, \citep{Z3}, \citep{J1}: 

\begin{align} \label{A7}
\pi_{\gamma}^{i} \left(\boldsymbol \beta_{\gamma}^{i}, \sigma \hspace{1mm} \vert \hspace{1mm}  g \right) = \sigma^{-1} \textrm{N}_{k_{\alpha}}\left(\boldsymbol \beta_{\gamma}^{i} \hspace{1mm} \vert \hspace{1mm} \textbf{0}, g_{\gamma}^{i} \sigma^2 (\textbf{X}_{\gamma}^{T}(\mathbb{I} - T^{-1}\mathbb{1}\mathbb{1}^{T})\textbf{X}_{\gamma})^{-1}\right).
\end{align}Under this prior, we have an analytical form for the Bayes factor:

\begin{align} \label{A8}
B_{\gamma0} &= \widetilde{B}_{\gamma0}^{1} \times \cdots \times  \widetilde{B}_{\gamma0}^{q}
\\
&= \Pi_{i=1}^{q} \frac{\left(1 +  g_{\gamma}^{i}\right)^{(T-k_{\gamma}-1)/2}}{\left(1 +  g_{\gamma}^{i}\frac{SSE_{\gamma}^{i}}{SSE_{0}^{i}}\right)^{(T+1)/2}},
\end{align}where $SSE_{\gamma}^{i}$ and $SSE_{0}^{i}$ are the sum of squared errors from the linear regression of column $\textbf{R}^{i}$ on covariates $\textbf{X}_{\gamma}$ and $k_{\gamma}$ is the number of covariates in model $M_{\gamma}$.  We allow the hyper parameter $g$ to vary across columns of $\textbf{R}$ and depend on the model, denoted by writing, $g_{\gamma}^{i}$.
\\
\\
We aim to explore the posterior of the model space, given our data:
\begin{align} \label{A9}
\textbf{P}\left(M_{\gamma} \hspace{1mm} \vert \hspace{1mm} \textbf{R} \right) = \frac{B_{\gamma0} \textbf{P}\left(M_{\gamma}\right)}{\Sigma_{\gamma} B_{\gamma0} \textbf{P}\left(M_{\gamma}\right)},
\end{align}where the denominator is a normalization factor.  In the spirit of traditional stochastic search variable selection \cite{OnSam}, we propose the following Gibbs sampler to sample this posterior.

\subsection{Gibbs Sampling Algorithm}

Once the parameters $\boldsymbol \beta_{\gamma}$ and $\sigma$ are integrated out, we know the form of the full conditional distributions for $\gamma_{i} \hspace{1mm} \vert \hspace{1mm} \gamma_{1}, \cdots, \gamma_{i-1}, \gamma_{i+1}, \cdots, \gamma_{p}$.  We sample from these distributions as follows:

\begin{enumerate}
\item Choose column $\textbf{R}^{i}$ and consider two models $\gamma{a}$ and $\gamma_{b}$ such that:
\begin{align*}
\gamma_{a} = (\gamma_{1}, \cdots, \gamma_{i-1}, 1, \gamma_{i+1}, \cdots, \gamma_{p})
\\
\gamma_{b} = (\gamma_{1}, \cdots, \gamma_{i-1}, 0, \gamma_{i+1}, \cdots, \gamma_{p})
\end{align*}

\item For each model, calculate $B_{a0}$ and $B_{b0}$ as defined by \ref{A8}.

\item Sample 
\begin{align*}
\gamma_{i} \hspace{1mm} \vert \hspace{1mm} \gamma_{1}, \cdots, \gamma_{i-1}, \gamma_{i+1}, \cdots, \gamma_{p} \sim Ber(p_{i})
\end{align*}

where

\begin{align*}
p_{i} = \frac{B_{a0} \textbf{P}\left(M_{\gamma_{a}}\right)}{B_{a0} \textbf{P}\left(M_{\gamma_{a}}\right) + B_{b0} \textbf{P}\left(M_{\gamma_{b}}\right)},
\end{align*}

\end{enumerate}

Using this algorithm, we visit the most likely ETF factor models given our set of target assets.  Under the model and prior specification, there are closed-form expressions for the posteriors of the model parameters $\beta_{\gamma}$ and $\sigma$.

\subsection{Hyper Parameter for the $g$-prior}

We use a local empirical Bayes to choose the hyper parameter for the $g$-prior in \ref{A7}.  Since we allow $g$ to be a function of the columns of $\textbf{R}$ as well as the model defined by $\gamma$, we calculate a separate $g$ for each univariate Bayes factor in \ref{A7} above.  An empirical Bayes estimate of $g$ maximizes the marginal likelihood and is constrained to be non-negative.  From \cite{Liang}, we have:

\begin{align}
\hat{g}_{\gamma}^{EB(i)} &= max\{F_{\gamma}^{i}-1,0\}
\\
F_{\gamma}^{i} &= \frac{R_{\gamma}^{2i} / k_{\gamma}}{(1-R_{\gamma}^{2i}) / (T - 1 - k_{\gamma})}.
\end{align}

For univariate stochastic search, the literature recommends choosing a fixed $g$ as the number of data points \cite{OnSam}.  However, the multivariate nature of our model induced by the multiple target assets makes this approach unreliable. Since each target asset has distinct statistical characteristics and correlations with the covariates, it is necessary to vary $g$ among different sampled models and target assets.  We find that this approach provides sufficiently stable estimation of the inclusion probabilities for the ETFs.

\newpage
\section{Simulation Study}

In this section of the appendix, we show results of applying our sampling algorithm to simulated data. Recall that we model the conditional, $R|X$, with parameters $\Psi$ and $\beta$ and the marginal, $X$, independently with parameters $\mu_{x}$ and $\Sigma_{x}$.  Using the posterior means of these parameters, we construct simulated target assets $R_{sim}$ and ETFs $X_{sim}$ under the data generating process:

\begin{align}
	X_{sim} &\sim N(\overline{\mu}_{x},\overline{\Sigma}_{x})
	\\
	R_{sim} &\sim \textrm{Matrix Normal}_{T,q}\left(X_{sim} \overline{\beta}, \hspace{1mm} \overline{\Psi}, \hspace{1mm} \mathbb{I}_{q \times q}\right),
\end{align}where the overlines represent the posterior means.  In \ref{sim1}, we show the true Sharpe ratio as well as its inferred value from our algorithm. The true value is calculated using \label{sim1} and the known moments of the data generating process for the simulated returns.  The Markov Chain Monte Carlo sampling does an excellent job at recovering the true Sharpe ratio as it is close to the posterior means from three separate simulated data sets.
\begin{figure}[H]
\centering
  \includegraphics[scale=.55]{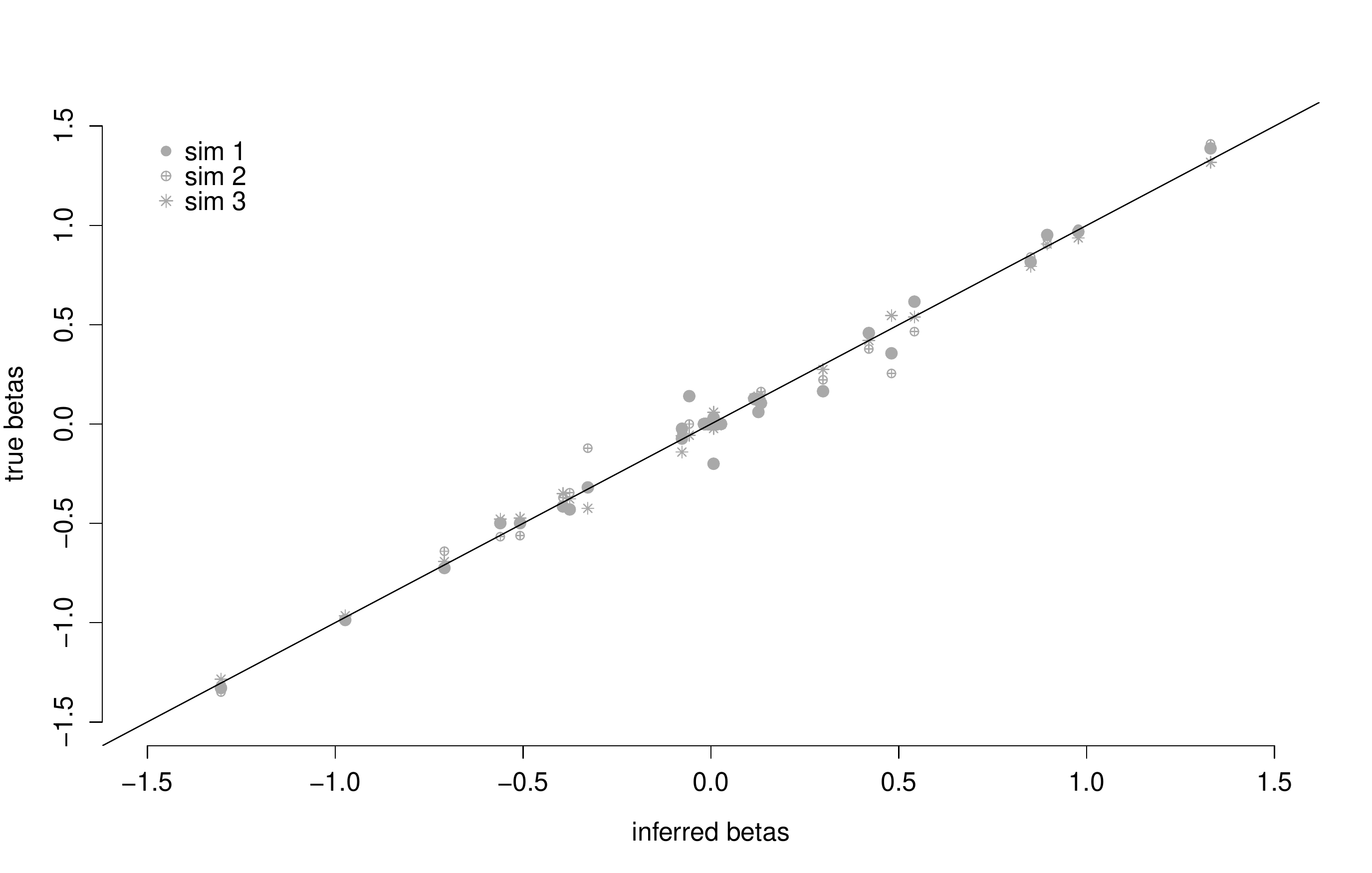}
  \caption{True versus inferred $\beta$'s for the three sets of simulated $R$ and $X$.}
  \label{betacompare}
\end{figure}

\begin{figure}
\centering
  \includegraphics[width=3in]{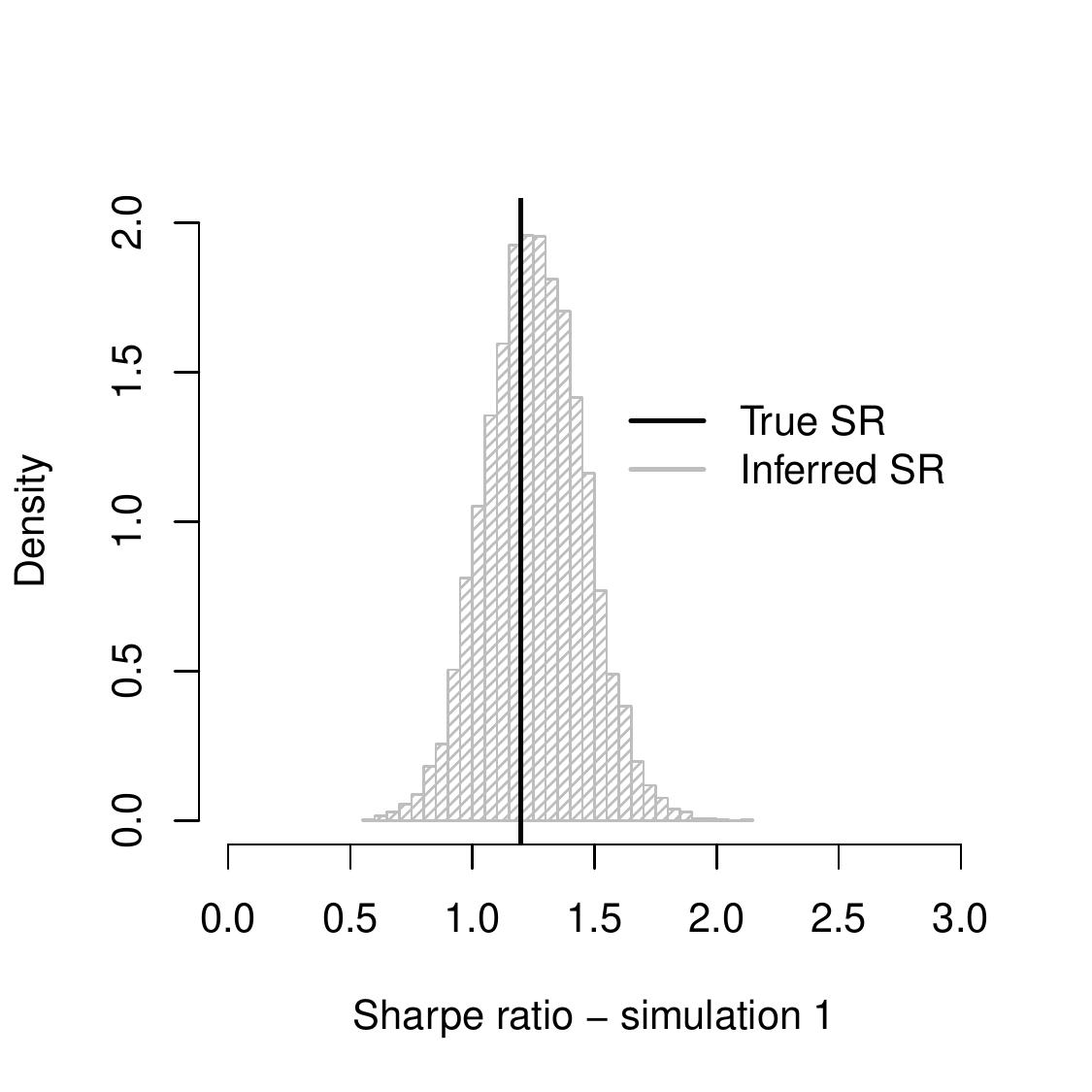}
  \includegraphics[width=3in]{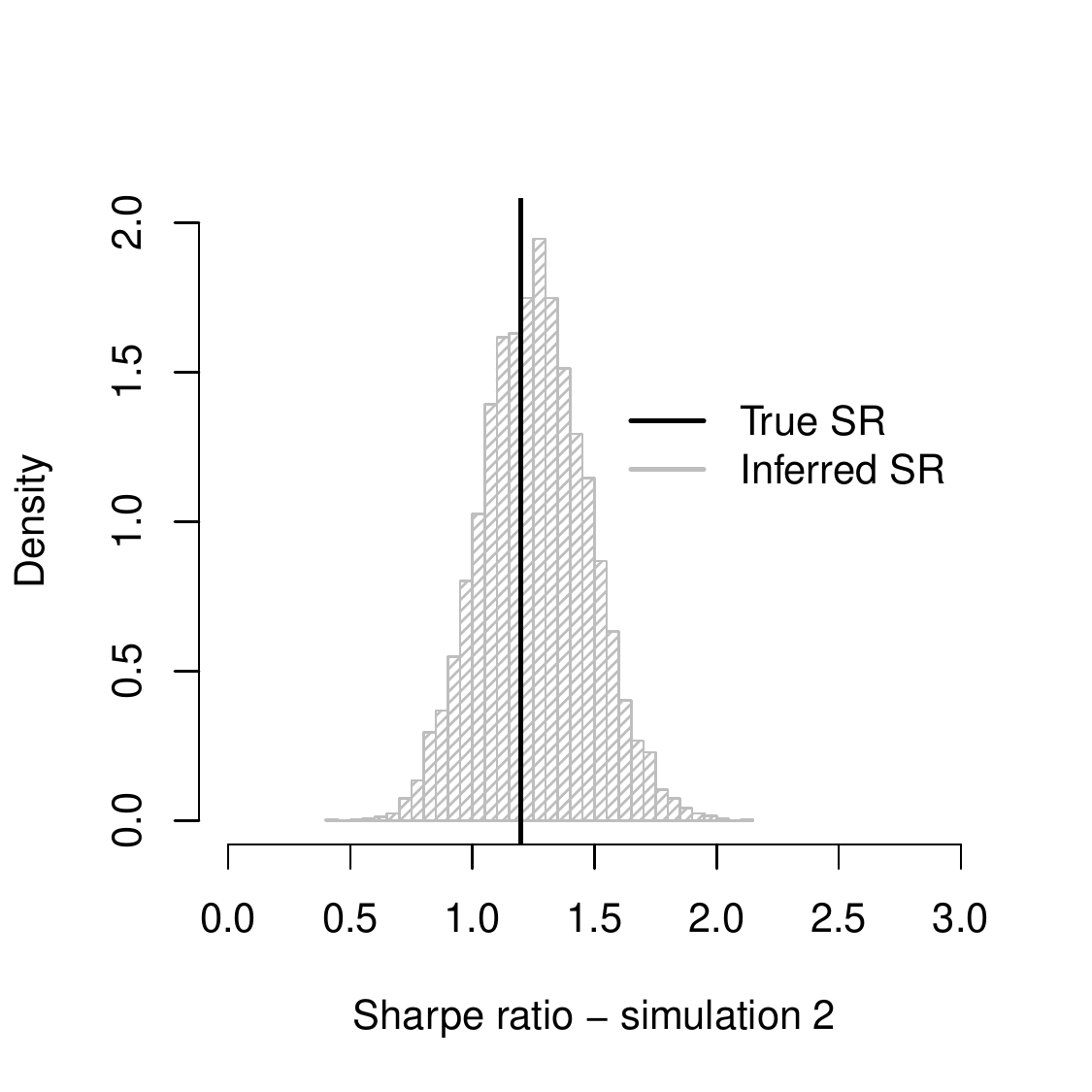}
  \includegraphics[width=3in]{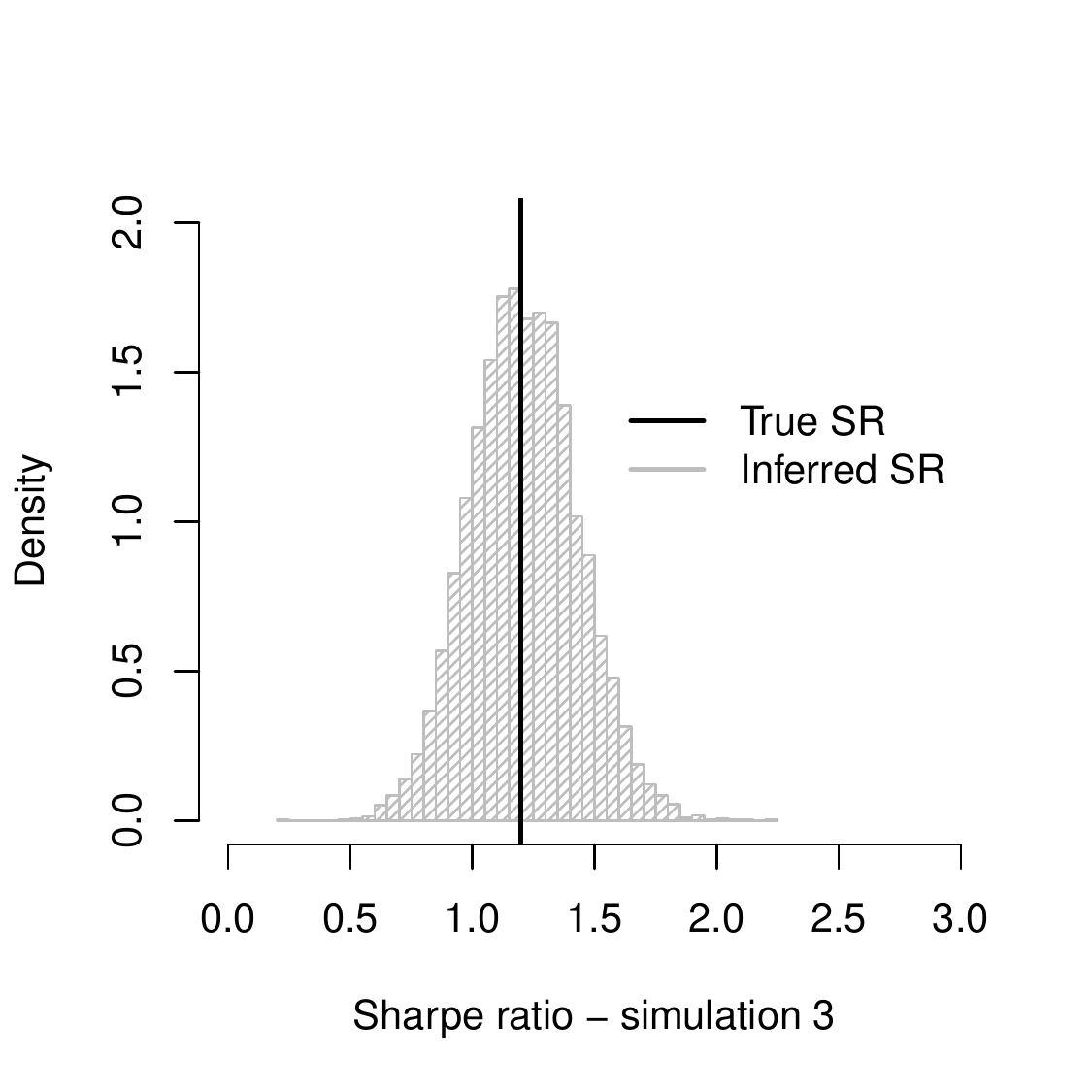}
  \caption{Posterior distribution of Sharpe ratios of the tangency portfolio for three simulate realizations of $R$ and $X$. The Sharpe ratio of the true tangency portfolio is shown as a vertical black line.}  \label{sim1}
\end{figure}

Additionally, we compare the posterior means of the $\beta$ coefficients with the true $\beta$'s used in each of the simulations.  The inferred $\beta$'s line up well with their true values as shown by their proximity to the 45 degree line in figure \ref{betacompare}.  Under reasonable data sets mimicking financial asset returns, our model sampling algorithm does well at recovering the parameters defining the data generating process.

\newpage
\section{Comparisons}
\subsection{Difference between conditional loss function and graphical lasso}
To demonstrate the difference between the conditional loss function and graphical lasso (glasso) approaches to the selection problem, consider a simple bivariate mean-zero model with one target asset and ETF.  Assume parameters $a,b,$ and $c$ have posterior means $\overline{a},\overline{b},$ and $\overline{c}$. 

\begin{equation}
\begin{split}
	\left(
	\begin{array}{c}
		r \\
		x
	\end{array}
	\right)
	&\sim N\left(\vec{0},\Sigma\right)
	\\
	\Sigma &= \left[
	\begin{array}{cc}
		a & c \\
		c & b
	\end{array}
	\right].
\end{split}	
\end{equation}  Analogous to the general setup, define the conditional model as:

\begin{equation}
	\begin{split}
		r \vert x &\sim N\left(\gamma x,d^{-1}\right).
	\end{split}
\end{equation} The goal is to achieve a parsimonious posterior summary of the off-diagonal element of $\Sigma$.  To achieve this, one may use the graphical lasso loss function discussed in the appendix of Hahn and Carvalho (2015) where the sparsification penalty only includes the off-diagonal element of our choice variable, which is the precision matrix, $\Gamma$:  

 \begin{equation}
 	\begin{split}
 		\Gamma = \left[
	\begin{array}{cc}
		\psi & g \\
		g & \kappa
	\end{array}
	\right].
 	\end{split}
 \end{equation}The glasso loss function from Hahn and Carvalho (2015) is: $\mathcal{L}(\Gamma) = \rho \norm{ \Gamma } - \log\det(\Gamma) + \text{tr}(\overline{\Sigma}\Gamma)$.  Note that $\rho$ is the parameter controlling the amount of penalization.  Only penalizing the dependence between $r$ and $x$, we simplify the loss function to:  

\begin{equation}
	\begin{split}
		\mathcal{L}_{\text{glasso}}(g,\psi,\kappa) = \rho| g | - \log(\psi\kappa - g^{2}) + (\overline{a}\psi + \overline{b}\kappa + 2\overline{c}g). \label{conlossex}
	\end{split}
\end{equation}The conditional loss function employed in the paper analogous with equation \ref{condlossfun} is:

\begin{equation}
	\begin{split}
		\mathcal{L}(\gamma) = \lambda | \gamma | -\frac{1}{2}\gamma^{2}\overline{b} + \gamma\overline{c}, \label{simconloss}
	\end{split}
\end{equation}where $\lambda$ is the penalization parameter.  A very important point involves the comparison of \ref{conlossex} and \ref{simconloss}. In glasso, $g$ is an element in the precision matrix.  Therefore, the implied covariance block for a choice of $g$ (using the 2x2 matrix inversion formula) is: $\gamma_{\text{glasso}}^{*} = -g\det\overline{\Sigma}$. In contrast, the conditional loss function choice variable, $\gamma$, is directly the coefficient matrix on $x$ in the conditional distribution, $r \vert x$.  Thus, comparisons of the two solutions will be made between $\gamma_{\text{glasso}}$ and $\gamma$.

\subsubsection{Conditional loss function optimum}
The first order conditions give the optimal action for the conditional loss function \ref{simconloss}, $\gamma^{*}(\lambda)$: 

\begin{equation}
	\begin{split}
		\gamma>0 &\implies \gamma^{*}(\lambda) = \frac{\overline{c}-\lambda}{\overline{b}}
		\\
		\gamma<0 &\implies \gamma^{*}(\lambda) = \frac{\overline{c}+\lambda}{\overline{b}},
	\end{split}
\end{equation}where we divide the action space into $\gamma$ positive and negative to account for the derivative of the absolute value in the penalty.

\subsubsection{Glasso loss function optimum} There are three actions for the glasso optimization.  Combining the first order conditions on $\psi$ and $\kappa$, we conclude that:

\begin{equation}
	\begin{split}
		\psi = \frac{\overline{b}}{\overline{a}}\kappa.
	\end{split}
\end{equation}Substituting this ratio back into the first order conditions for $\psi$ and $\kappa$ and solving the resulting quadratic equation, we obtain $\psi$ and $\kappa$ as functions of the parameters and $\gamma^{2}$:

\begin{equation}
	\begin{split}
		\kappa = \frac{1}{2\overline{b}}\left(1 + \sqrt{1+4\overline{a}\overline{b}g^{2}}\right), \hspace{5mm} \psi &= \frac{1}{2\overline{a}}\left(1 + \sqrt{1+4\overline{a}\overline{b}g^{2}}\right). \label{kandp}
	\end{split}
\end{equation}We take the positive roots to ensure the diagonal elements of our action are positive.  This is necessary since glasso seeks a positive definite matrix.  The first order condition for $g$ when $g<0$ implies:

\begin{equation}
	\begin{split}
		-\rho + \frac{2g}{\psi\kappa - g^{2}} + 2\overline{c} = 0.\label{FOC}
	\end{split}
\end{equation}The first order condition for the case of $g>0$ is the same, but with a positive sign on $\rho$.  Substituting \ref{kandp} into \ref{FOC}, we obtain the optimal action, $g^{*}(\rho)$:  

\begin{equation}
	\begin{split}
		g>0 &\implies g^{*}(\rho) = \frac{\frac{1}{2}\rho - \overline{c}}{\det{\overline{\Sigma}} + \overline{c}\rho - \frac{1}{4}\rho^{2}}
		\\
		g<0 &\implies g^{*}(\rho) = \frac{-\frac{1}{2}\rho-\overline{c}}{\det{\overline{\Sigma}} - \overline{c}\rho - \frac{1}{4}\rho^{2}},
	\end{split}
\end{equation}where:

\begin{equation}
\begin{split}
	\overline{\Sigma} &= \left[
	\begin{array}{cc}
		\overline{a} & \overline{c} \\
		\overline{c} & \overline{b}
	\end{array}
	\right].
\end{split}	
\end{equation}

The unpenalized solutions for our conditional loss function and the graphical lasso are:

\begin{equation}
\begin{split}
	\gamma^{*}(0) &= \frac{\overline{c}}{\overline{b}}
	\\
	\gamma_{\text{glasso}}^{*}(0) &= -g^{*}(0)\det\overline{\Sigma} = \overline{c}
\end{split}	
\end{equation}

\subsubsection{Numerical demonstration}

Given the derived solutions for the conditional loss function and graphical lasso optimizations, we provide a numerical example of their difference.  We set $\overline{a}=12$, $\overline{b}=1, \text{ and } \overline{c}=3$. Setting $\overline{b}=1$ guarantees that the unpenalized solutions: $\gamma^{*}(0)$ and $\gamma_{\text{glasso}}^{*}(0)$ will be equal. Further, since $\Sigma$ must be positive definite, we have that its determinant, $\overline{a}\overline{b} - \overline{c}^{2} = 3$, is positive. 

\begin{figure}[H]
\centering
  \includegraphics[scale=.55]{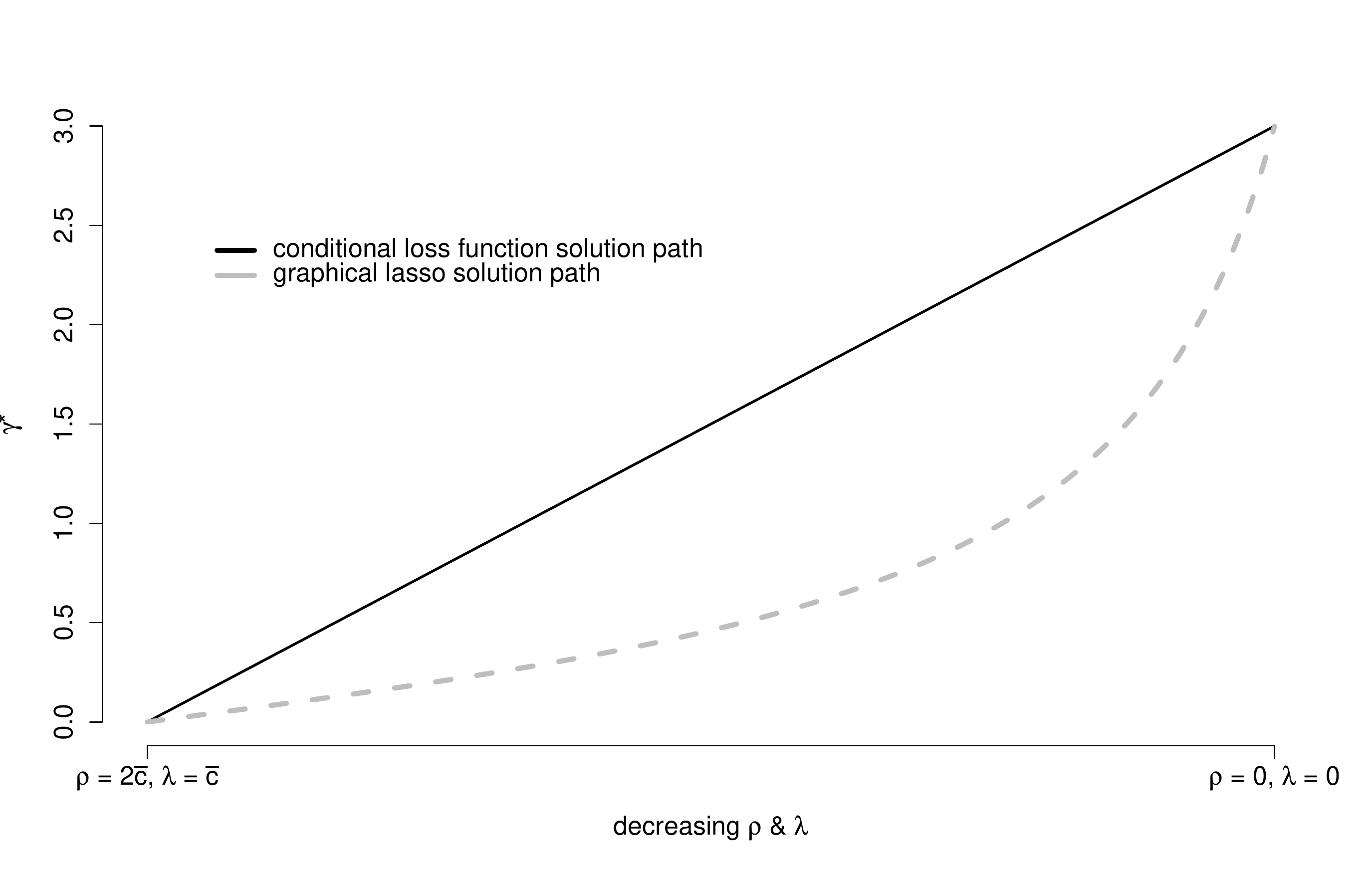}
  \caption{$\overline{a}=12$}
  \label{compgraph}
\end{figure}

 Figure \ref{compgraph} displays how the optimal solutions for the conditional loss function and graphical lasso change with their penalty parameters - known as the solution paths.  For simplicity, we plot both penalty parameters on the same axis.  The left part x-axis is the beginning of the solution path where the penalties are large enough to send the solutions to zero.  This occurs when $\rho=2\overline{c}$ and $\lambda=\overline{c}$.  The right part of the x-axis shows the unpenalized solutions, and they are designed to be equal in our example.
 
\begin{figure}[H]
\centering
  \includegraphics[scale=.55]{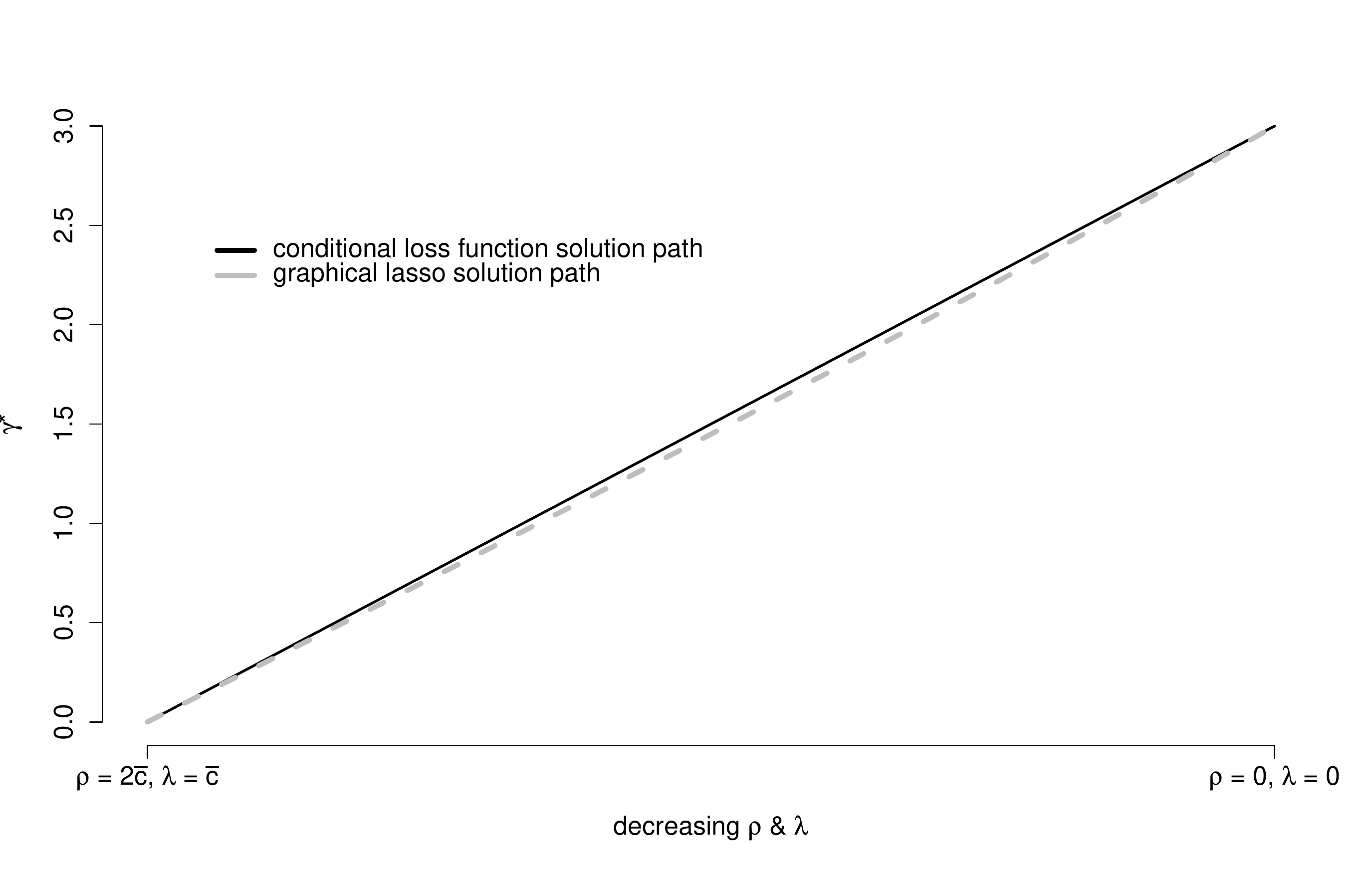}
  \caption{$\overline{a}=200$}
  \label{compgraph1}
\end{figure}

We see in figure \ref{compgraph} that the solution paths are very different.  The graphical lasso solution depends nonlinearly on its penalty parameter, $\rho$.  The concavity of its solution path can be increased by decreasing $\overline{a}$ towards its constrained value required by $\det\overline{\Sigma} > 0$, necessary for positive definiteness.  When $\overline{a}=200$ as in figure \ref{compgraph1}, the graphical lasso solution path becomes much more linear, and the two paths begin to coincide (they are, however, not the same due to the different penalty scales and a choice of $\overline{b}$ other than 1 would affect the slopes).  This occurs when the $\det\overline{\Sigma}$ dominates the numerator and denominator terms in $\gamma_{\text{glasso}}^{*}(\rho) = -g^{*}(\rho)\det\overline{\Sigma}$.  Intuitively, this can also be understood through a correlation argument.  As $\overline{a}$ gets large, the correlation between $r$ and $x$ squared: $\frac{\overline{c}^{2}}{\overline{a}\overline{b}}$ goes to zero.  This increased ``independence" between $r$ and $x$ results in the penalized graphical lasso objective function becoming exactly the conditional loss objective function.  In fact, one could substitute the optimal solutions for $\kappa$ and $\psi$ (\ref{kandp}) into the glasso objective function (\ref{conlossex}) and Taylor expand about $g$ (since $g^{*}(\rho)$ is small when $\overline{a}$ is large) to directly see the similarity between the two optimizations.  

\newpage
\bibliographystyle{apalike.bst}
\bibliography{ETFTangencyPortfolio}

\end{document}